*Correspondence:
Vadim Volkov,
Faculty of Life Sciences, School of
Human Sciences, London
Metropolitan University, 166-220
Holloway Road, London N7 8DB, UK
vadim.s.volkov@gmail.com

# Quantitative description of ion transport via plasma membrane of yeast and small cells

Vadim Volkov*

Faculty of Life Sciences, School of Human Sciences, London Metropolitan University, London, UK

Modeling of ion transport via plasma membrane needs identification and quantitative understanding of the involved processes. Brief characterization of main ion transport systems of a yeast cell (Pma1, Ena1, TOK1, Nha1, Trk1, Trk2, non-selective cation conductance) and determining the exact number of molecules of each transporter per a typical cell allow us to predict the corresponding ion flows. In this review a comparison of ion transport in small yeast cell and several animal cell types is provided. The importance of cell volume to surface ratio is emphasized. The role of cell wall and lipid rafts is discussed in respect to required increase in spatial and temporary resolution of measurements. Conclusions are formulated to describe specific features of ion transport in a yeast cell. Potential directions of future research are outlined based on the assumptions.

Keywords: ion transport, yeast cell, erythrocyte, excitable membrane, lipid rafts, cell wall, salinity tolerance, systems biology

## Introduction

The fungus *Saccharomyces cerevisiae* (Phylum Ascomycota) is a well-known baker's yeast. It is a small unicellular organism (**Figure 1**), which can grow in a wide range of pH, osmolality and various ion compositions of surrounding media. Yeast cells are among the best studied unicellular eukaryotic organisms with small sequenced genome, large available collections of mutants in specific genes, high growth rate in nutrient media. They are easy for genetic and molecular biological manipulation. Essential volume of accumulated knowledge about yeast facilitates further research in the area.

Yeast cells are widely used in the food industry, for baking and for brewing, for making wine and spirits. More recent and advanced applications include biotechnology, chemical industry and pharmacology where yeast cells are producing pharmaceutical and nutraceutical ingredients, commodity chemicals, biofuels and also heterologous proteins including different enzymes from other eukaryotic organisms. The commercial scale of production is achieved for the novel applications based on progress in synthetic biology and metabolic engineering (e.g., reviewed in Borodina and Nielsen, 2014). Yeast cells are invaluable for applications in biomedical research. Heterologous expression of mammalian proteins, especially membrane ones in yeast is an important means to understand their properties. Eukaryotic yeast cells with specific mutant phenotypes could be rescued after expressing homologous or complementing proteins from the other organisms, thus giving indications about the functions and interactions of the proteins. Amino acid mutations and substitutions within the proteins of interest allow detailed analysis of their structure and protein domains. Yeast two-hybrid screening is a technique in molecular biology to understand protein-protein interactions (Fields and Song, 1989; reviewed in Brückner et al., 2009); modifications of the method include split-ubiquitin system (Stagljar et al., 1998; reviewed in Thaminy et al., 2004) and several others for interacting membrane proteins *in vivo*. Yeast expression system helped to understand elements of calcium signaling in eukaryotic cells (reviewed in Ton and Rao, 2004). Yeast cells served for functional expression and characterization in more detail protein domains of an inward-rectifying mammalian $K^+$channel mKir2.1 (Hasenbrink et al., 2005, 2007; Kolacna et al., 2005), rat neuronal potassium channel rEAG1 (Schwarzer et al., 2008), human estrogen alpha and beta receptors (Sievernich et al., 2004; Hasenbrink et al., 2006; Widschwendter et al., 2009), human receptors of the Hedgehog pathway (Joubert et al., 2010) and the other heterologous membrane proteins for functional and structural studies. The methods and applications are essentially based on the present and growing knowledge of yeast cells, their genetics, molecular biology and physiology.

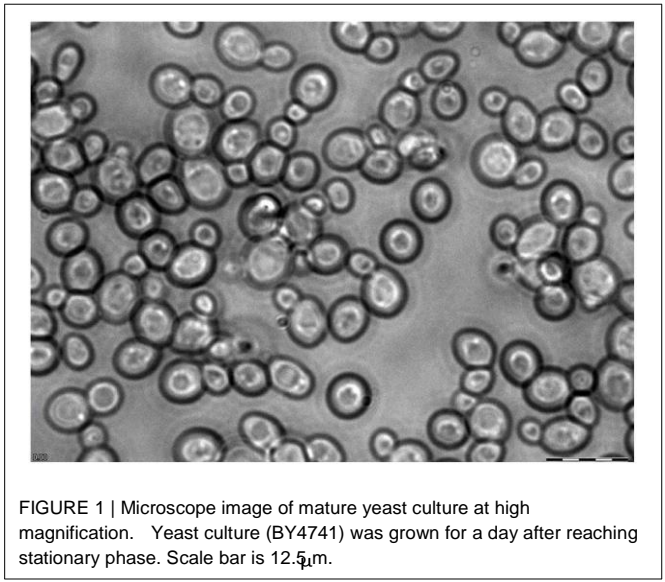

FIGURE 1 | **Microscope image of mature yeast culture at high magnification.** Yeast culture (BY4741) was grown for a day after reaching stationary phase. Scale bar is 12.5 μm.

Ion homeostasis is important for yeast growth (cell volume increases mainly by water uptake according to osmotic gradient against cell wall mechanical pressure) and also for the function of enzymes. Some proteins (e.g., yeast *HAL2* nucleotidase) may change conformation and lose activity under increased concentrations of $Na^+$ (Murguía et al., 1996; reviewed in Serrano, 1996; Albert et al., 2000). Understanding, describing and modeling ion transport is important for optimizing and improving growth conditions for yeast culture.

Initial assumptions for modeling seem oversimplified for a biologist; however, they are required for the basic biophysical description of the processes. The cell is considered to be a homogeneous spherical body consisting of viscous cytoplasm containing several ion species and surrounded by a lipid membrane. The lipid membrane contains a large number of incorporated proteins (ion pumps, channels, and transporters), which make pathways for selective and non-selective transport of ions. The cell is further surrounded by the cell wall.

Inner cell structures are present (e.g., nucleus, ATP producing mitochondria, clusters of so called lipid rafts within the plasma membrane, possible vacuolization and existing intracellular compartments etc.) and will be mentioned if necessary.

The numeric parameters of a yeast cell are—cell volume, membrane surface area, ion concentrations within and outside of the cell, yeast cell electric membrane potential, characteristics and number of ion transport systems of a yeast cell and also mechanical properties (elastic and plastic elasticity) of the cell wall. The presence of cell wall is a similarity between yeast, plant, algal and most of prokaryotic cells, while making them distinct from most of animal cells. Stretching cell wall balances hydrostatic turgor pressure, which is developed from the interior of the cell due to difference in osmotic pressures inside and outside of the cell. Positive turgor pressure is caused by water fluxes into the cell following higher concentration of osmotically active compounds inside. Ion gradients and partially the higher osmotic pressure are created by the concerted activity of ion pumps, channels and transporters, which also keep stable or ensure perturbed for signaling ion concentrations; ion transport systems are also responsible for negative membrane potential.

Exploring yeast with small size of their cells (several μm or around 10 wavelengths of red light) breaks trivial everyday experience about the world resembling to what is observed in microbiology (Beveridge, 1988) and cell biology (AlbrechtBuehler, 1990), hence requires special knowledge and equipment.

## Quantitative Characteristics of Yeast Cells

Assuming an average diameter of yeast cell of about 6 μm and approximating the cell as a spherical body (**Figure 1**), we can calculate the **volume of the yeast cell** according to formula linking volume to diameter of sphere:

$$4/3 * \pi * (6/2)^3 * 10^{-18} m^3 = 4/3 * \pi * 27 * 10^{-15} L \approx 100 \text{ fL} = 0.1 \text{pL}. \tag{1}$$

For comparison, the volume of simpler prokaryotic microbe *Escherichia coli* is about 1,5-4,4 fL (e.g., Volkmer and Heinemann, 2011), the volume of a mammalian spermatozoon is about 20–30 fL (Curry et al., 1996). The volume of a human erythrocyte is also about 100 fL (Jay, 1975), but the volume of typical mammalian cardiomyocytes is 300 times larger (Satoh et al., 1996). The volume of giant squid axon is $10^8$ times larger (5 cm long and 500 μm in diameter) (Lecar et al., 1967), the volume of barley leaf protoplasts is at least 100 times larger (Volkov et al., 2009).

It is important to mention that the basics of membrane ion transport were initially formulated in the outstanding works of Nobel Prize laureates Hodgkin and Huxley for large cells (squid axon) (Hodgkin and Huxley, 1952), so the principles for small cells with much higher surface/volume ratio may need to be scaled and amended. The surface/volume ratio is the surface per unit of volume; it consequently determines ion fluxes for the unit of volume and, *vice versa*, the potential metabolic activity per unit of surface (**Figure 2**). The surface/volume ratio is 125 times higher for a yeast cell than for a squid axon; this poses questions about potential principal differences in ion transport systems.

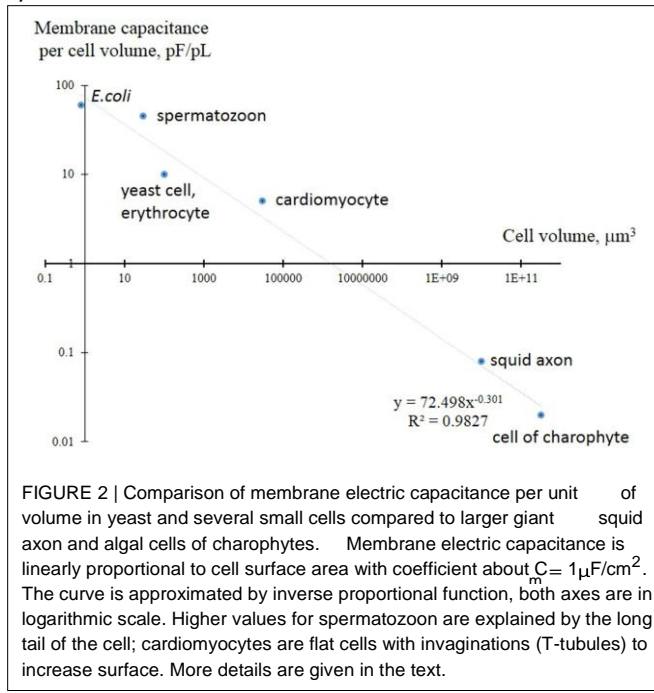

FIGURE 2 | Comparison of membrane electric capacitance per unit of volume in yeast and several small cells compared to larger giant squid axon and algal cells of charophytes. Membrane electric capacitance is linearly proportional to cell surface area with coefficient about $C_m = 1 \mu F/cm^2$. The curve is approximated by inverse proportional function, both axes are in logarithmic scale. Higher values for spermatozoon are explained by the long tail of the cell; cardiomyocytes are flat cells with invaginations (T-tubules) to increase surface. More details are given in the text.

**The cell surface area** for a yeast cell is:

$$4*\pi*9 \ \mu m^2 \approx 100 \ \mu m^2. \qquad (2)$$

This value of surface area of biological membrane corresponds to electric capacitance 1 pF, since 100 $\mu m^2$ make up about 1 pF (the specific electric capacitance of biological membranes $C_m$ is about $C_m = 1\mu F/cm^2$) (Hille, 2001). Similar results for capacitance are indeed found in patch clamp experiments (0.5– 0.7 pF for yeast spheroplasts with diameters of 4–5 μm) (Roberts et al., 1999). This means that the electric charge q transfer of N = 600,000 monovalent cations (e.g., potassium, proton or sodium) out of cell will cause the change of voltage V from 0 to −100 mV according to the definition of electric capacitance C (q = C∗V) and taking into account the elementary charge $1.6*10^{-19}$ coulombs (equal to the electric charge of a monovalent cation; hence we have about $6.2*10^{18}$ monovalent cations per coulomb):

$$N = (10^{-12}F*0.1 \ V)/1.6*10^{-19} \text{coulombs} \approx 6*10^5. \qquad (3)$$

This calculated number of cations is about 0.01% of $K^+$ ions in a yeast cell ($6*10^9$ ions per cell for 100 mM $K^+$, see below), however 100 times higher than the estimated number of free protons in the yeast cell (without a vacuole and without considering any pH buffering in the cell). **pH of the cytoplasm** of a yeast cell is about 7.0, which means $10^{-7}$ M $H^+$/L, cell volume is 100 fL, therefore:

$$10^{-7} \ M \ H^+/L*100*10^{-15} \ L*N_a \approx 10^{-20} \ \text{mole} \ H^+*6.02*10^{23}/\text{mole}$$

$$\approx 6,000 \ \text{protons/cell}, \qquad (4)$$

where $N_a = 6.02*10^{23}$/mole is the Avogadro constant corresponding to the number of ions/molecules per mole of a chemical compound/substance.

Similar assumptions and calculations for the number of ions and voltage changes can be found also in Kahm (2011), for example, where several differential equations have been proposed and based on theoretical considerations to build a model of potassium homeostasis in *Saccharomyces cerevisiae*.

**Ion concentrations** in yeast cells were measured by several methods and gave readings about 100–150 mM for $K^+$ (Mulet and Serrano, 2002; Jennings and Cui, 2008; Zahrádka and Sychrova, 2012). However, concentrations of 50–300 mM of $K^+$ have been reported (depending on growth phase, external potassium etc., reviewed in Ariño et al., 2010; Kahm, 2011). Under low (0.1 mM) external $K^+$ at the background of 100– 300 mM NaCl internal $K^+$ concentration below 5 mM has been measured (Alemán et al., 2014). $Na^+$ depends more on the external concentration and may vary from a few mM to over 100 mM (García et al., 1997; Kolacna et al., 2005). For example, addition of external 1 M NaCl increased internal $Na^+$ concentration linearly from 0 to 150 mM after 100 min (García et al., 1997). Chloride concentration is relatively low and stable, about 0.1–1 mM, indicating that chloride is also a homeostatically controlled abundant ion in the cell (Jennings and Cui, 2008). Calcium concentration in yeast and eukaryotic cells is very low, 50–200 nM. This ion is essential for signaling, so specific pumps and transporters exist to ensure calcium homeostasis and signaling (e.g., Cui and Kaandorp, 2006).

These concentrations correspond to relatively small numbers of ions per cell. Even the number of $K^+$ ions is amenable for simple modeling using modern software and computers. Instead and in addition to using thermodynamic approach, location and properties of each ion could be potentially digitized and analyzed.

$$100 \ \text{mM} \ K^+ \ \text{in} \ 100 \ \text{fL gives} \ N^*_a 100 \ \text{mM}*100 \ \text{fL}$$

$$\approx 6.02*10^{23}/\text{mole}*100 \ \text{mmole/L}*100 \ 10^{-15} \ L$$

$$\approx 6*10^9 \ \text{ions of} \ K^+/\text{cell}, \qquad (5)$$

where $N_a = 6.02 * 10^{23}$/mole is the Avogadro constant.

This number of ions is equivalent to an electrical current of $1.6*10^{-19}$ coulombs/ion $*6*10^9$ ions $\approx 10^{-9}$ A*seconds = 10 seconds * 100 pA (elementary charge of cation $1.6*10^{-19}$ coulombs multiplied by the number of ions and converted to Amperes; will be discussed later when considering ion currents via ion channels and transporters).

Direct measurements of **membrane potential** using glass microelectrodes is the only direct method available to measure membrane potential. However, for small cells of *Saccharomyces cerevisiae* the method may be inaccurate and can underestimate membrane potential considering that (1) membrane potential could be changed by the proton pump Pma1 within seconds (see discussion below) and (2) microelectrodes may have an effect on the measured values, for example, they cause KCl leak and mechanical stress (e.g., Blatt and Slayman, 1983). Therefore, indirect methods and comparative results for different yeast species and under several conditions are very important.

Microelectrode technique records membrane potentials of around −70 to −45 mV for *S. cerevisiae* (reviewed in BorstPauwels, 1981) and similar or lower values for other yeast species (summarized in **Table 1**).

response to addition of 10 mM KCl (Peña et al., 2010). The effect of depolarization by KCl was observed in the presence of 20 mM glucose suggesting the implication of proton pump Pma1 in membrane energization (Peña et al., 2010). Recordings with 3,3′-dipropylthiacarbocyanine demonstrated depolarization for aerobic yeast *Rhodotorula glutinis*: the cells depolarized by 25 mV in 25 mM KCl, by 60 mV in 100 mM KCl and by 60 mV, when external pH changed from 6 to 3 (Plášek et al., 2012). In the presence of 100 mM glucose cells of *S. cerevisiae* depolarized by 30 mV in 100 mM KCl, by 20 mM in 100 mM NaCl and by 40 mV under pH decrease from 6 to 3 (Plášek et al., 2013). In the absence of glucose depolarization was larger for pH change and 100 mM KCl though smaller for 100 mM NaCl (Plášek et al., 2013).

This shows that indirectly measured membrane potential values are similar (Felle et al., 1980; Lichtenberg et al., 1988) or more negative (Vacata et al., 1981) than when measured directly by microelectrodes due to a variety of reasons, e.g., impalement problems and leak from microelectrodes, or nonspecific absorption of cations.

TABLE 1 | Values of membrane potential of yeast and several fungal and bacterial cells recorded by microelectrodes.

| Species | Membrane potential values, mv | References | Notes |
|---|---|---|---|
| *Saccharomyces cerevisiae* | Around −70 to −45 | Reviewed in Borst-Pauwels, 1981 | Values might be not realistic, fast decay in some experiments |
| *Endomyces magnusii*, (size of cells is about 15×30 μm) | (1) Around −40 (2) Around −150 to −120 | (1) Vacata et al., 1981 (2) Bakker et al., 1986 | (1) Measured in artificial pond water, pH is not indicated, over 1.5 mM K$^+$, 0.5 mM Na$^+$, 0.2 mM Ca$^{2+}$ (2) −124 mV at pH 4.5 (0.1 mM KCl), preimpalement value estimated around −190 mV; −146 mV at pH 7.1 (0.1 mM KCl) and preimpalement value estimated around −275 mV |
| *Pichia humboldtii* (cell size is about 12–15 μm in diameter) | Around −90 to −50 | Höfer and Novacky, 1986; Lichtenberg et al., 1988 | 100 mM KCl depolarized membrane of *Pichia humboldtii* from −48 to −21 mV, while 100 mM NaCl to −37 mV and 100 mM LiCl by 3 mV only; membrane potential depended on external pH increasing from −60 to −30 mV under external pH change from 6 to 8 or from 6 to 4 Höfer and Novacky, 1986 |
| *Neurospora. crassa* (a) hyphae diameter over 10–15 μm or (b) spherical cells with diameter about 15–20 μm) | Below −300 to −120 | (a) Slayman and Slayman, 1962; (b) Blatt and Slayman, 1983, 1987 | Depended on concentration of external K$^+$ and internal pH |
| Giant *E. coli* cells (about 5 μm in diameter) | Below −140 | Felle et al., 1980 | −100 mV at pH 5.5 and −142 mV at pH 8.0 |

Indirect methods of membrane potential measurements are based on steady-state distribution of lipophilic cations and require further simple calculations; tetra[$^3$H]—phenylphosphonium (TPP$^+$) and tri[$^3$H]phenylmethylphosphonium (TPMP$^+$) are used to measure membrane potential in yeast cells. With this method values around −120 mV to −50 mV were estimated (Vacata et al., 1981). Measurements using 3,3′-dipropylthiacarbocyanine gave estimates around −160 to −150 mV for *S. cerevisiae*; the membrane potential depolarized within 10 s by 25–30 mV in

## Ion Transport Systems of Yeast Cells

Ion transport systems have been well studied for yeast cells within the last 30–40 years following the rise of molecular biology, electrophysiology and complete sequencing of the yeast genome. The new methods complement traditional experiments with yeast mutants and ion accumulation by yeast cells to elucidate the mechanisms of transport and role of specific proteins. The main ion transport systems include ion pumps, several transporters, one potassium channel and a non-selective cation current, which was recorded in electrophysiological experiments (**Figure 3**).

## Ion Pumps

The most abundant protein in the yeast plasma membrane is the **proton pump Pma1** (P-type H$^+$-ATPase), which pumps protons out of the cytoplasm and shifts the membrane potential to more negative values. Pma1 transports one H$^+$ per ATP molecule (Ambesi et al., 2000) and makes up about 15–20% of membrane proteins (Eraso et al., 1987). Protein expression analysis (Ghaemmaghami et al., 2003) estimates the amount of Pma1 per cell to be 1,260,000 Pma1 molecules/cell. This is equivalent to 12,600 molecules/μm$^2$ and equates to one Pma1 molecule per 10∗10 nm$^2$ of membrane surface. The value is above the feasible limits, since the linear size of proton ATPases is large and, for example, the cross-section of similar H$^+$-ATPase from the plasma membrane of *Neurospora crassa* is nearly 10∗10 nm$^2$ (Auer et al., 1998). Assuming that only 10% of expressed Pma1 is present in the plasma membrane (many molecules could be in secretory vesicles in cytoplasm) we get 130,000 molecules per plasma membrane. This number coincides better with electron microphotographs of yeast plasma membranes labeled with antibodies for Pma1 (Permyakov et al., 2012).

One Pma1 molecule transports 20–100 H$^+$ per second (Serrano, 1988). The conclusion is confirmed by measurements of ATP hydrolysis activities (e.g., Perlin et al., 1989): assuming transport of one H$^+$ per one reaction of ATP hydrolysis and knowing ATPase activities for reconstituted mutant enzymes in μmol P$_i$ mg$^{-1}$ min$^{-1}$ we can estimate the number of transported H$^+$/(Pma1 molecule∗second). The value is similar to animal Na$^+$/K$^+$-ATPase with a turnover of 160 (Skou, 1998). Taking the lower of these values it is simple to estimate the possible number of protons extruded by Pma1 within one second per yeast cell:

$$20 H^+/(second*pump\ molecule)*130,000 pump\ molecules/cell = 2,600,000\ H^+/(second*cell). \quad (6)$$

This value exceeds by a factor of four the number of positive charges required to shift the membrane potential of the cell by −100 mV in one second (Equation 3). Obviously, the situation is more complex since one must also consider thermodynamics of transport and ATP hydrolysis. With a cytoplasmic pH similar to ambient medium Pma1 will not be able to pump protons against a membrane potential lower than -400 to -450 mV due to limitation by energy of ATP hydrolysis: −30 kJ/mole to be translated to the energy of transporting an elementary charge against the membrane voltage and would correspond to transport against −300 mV, slightly more negative hydrolysis energy could be achieved and result in more negative voltages (e.g., Bond and Russel, 1998). Indeed, the most negative membrane potentials recorded in fungal/autotrophic cells are around −300 to −350 mV for *Neurospora crassa* with equilibrium potential up to −450 mV (Blatt and Slayman, 1987).

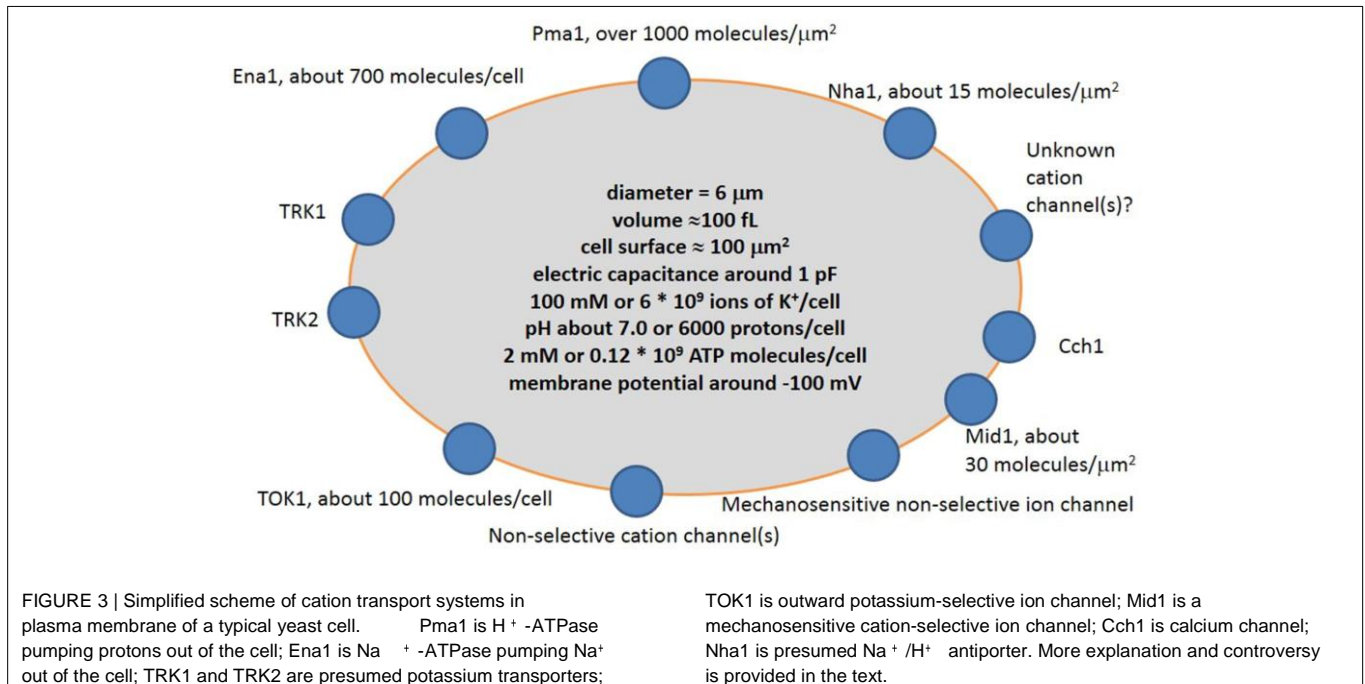

FIGURE 3 | Simplified scheme of cation transport systems in plasma membrane of a typical yeast cell. Pma1 is H$^+$-ATPase pumping protons out of the cell; Ena1 is Na$^+$-ATPase pumping Na$^+$ out of the cell; TRK1 and TRK2 are presumed potassium transporters; TOK1 is outward potassium-selective ion channel; Mid1 is a mechanosensitive cation-selective ion channel; Cch1 is calcium channel; Nha1 is presumed Na$^+$/H$^+$ antiporter. More explanation and controversy is provided in the text.

These simple considerations show that the most important factors for membrane potential and driving force for translocation of protons (1μH$^+$) in yeast cell are:

(1) ATP availability and cell metabolism;
(2) Pma1 regulation, not the potential maximal Pma1 activity;

(3) futile cycles of ion transport and inward electrogenic ion transport.

ATP concentration in yeast cells is about 2 mM in the presence of glucose (measured by aptamer-based ATP nanosensor) (Ozalp et al., 2010) or around: 250–350 attomole ATP per a cell (Ashe et al., 2000) estimated to 2.5–3.5 mM; 8.4 ± 0.4 µmole ATP/3*10$^7$ cells (Ullah et al., 2013) recalculating to 2.8 mM. It is similar to the ATP concentration in bacteria, for example 3 mM ATP was measured in *Streptococcus bovis* (Bond and Russel, 1998). The total number of ATP molecules will be in that case:

$$2 \text{ mM ATP in 100 fL gives } N^*_a \, 2 \text{ mM}^*100 \text{ fL}$$

$$\approx 6.02^*10^{23}/\text{mole}^*2 \text{ mmole/L}^*100^*10^{-15}\text{L}$$

$$\approx 1.2^*10^8 \text{ ATP molecules/cell, where}$$

$N_a = 6.02^*10^{23}$/mole is the Avogadro constant.

This is at least 50 times higher than the potential minimal requirement of ATP consumed by Pma1 per a second (Equation 6); therefore ATP is likely not a limiting factor immediately linking metabolic activity of yeast cell with fast membrane transport processes (timescale of seconds). If the highest number of Pma1 in the membrane ($1.3^*10^6$ molecules) is assumed (not likely due to above mentioned steric calculations for the molecules) and the turnover of 100 H$^+$/(Pma1 molecule*second) is taken, the amount of ATP is still sufficient for a second. There is a reasonable time lapse of 1–50 s (buffered by internal existing concentration of ATP) between (1) the proton transport and plasma membrane energization by Pma1 and (2) cell metabolism and ATP synthesis. It is worth to include for consideration ATP/ADP ratio, energy charge of yeast cells, concentrations of free phosphate and ADP. The parameters influence the change of the Gibbs free energy 1G for ATP hydrolysis (e.g., Bond and Russel, 1998) and also regulate many biochemical and physiological reactions in cells (e.g., Pradet and Raymond, 1983; Wilson et al., 1996; Gout et al., 2014) linking ion transport and cell metabolism. Total concentration of adenine nucleotides in cytoplasm of yeast cell was measured around 3–5 mM (Brindle et al., 1990; Nielsen et al., 2010), while ATP/ADP ratio varied from about 5 in high glucose to 0.3 after 15 min in low glucose, ATP/AMP ratio dropped from over 23 to 0.1 under the switch from high to low glucose (Wilson et al., 1996). Intracellular phosphate was similar to adenine nucleotides, 2.5–5.2 mM (Brindle et al., 1990). Complex and not always predictable changes of cytoplasmic nucleotide concentrations follow cell metabolism. For example, addition of 100 mM glucose at the background of 2% trehalose five-fold decreased ATP concentration less than in 30 s (Loret et al., 2007). Interestingly and unexpectedly, the reduction of ATP was due to increase of IMP and inosine without essential change of ADP and AMP; concentrations of adenine nucleotides recovered to initial levels in about 30 min (Loret et al., 2007). It is a complicated network of transcription factors, protein kinases and the other regulatory proteins and factors involved in nutritional, energy and metabolic control in yeast cells (e.g., Lee et al., 2008; Broach, 2012; Österlund et al., 2013). Pma1 is phosphorylated by several protein kinases and changes affinity to ATP in response to glucose metabolism; it is a way to modulate membrane potential (Goossens et al., 2000). Regulation of Pma1 involves the other numerous proteins and networks of events and is the subject of specific reviews (e.g., discussed in Ariño et al., 2010 and referenced therein; Babu et al., 2012), therefore it will not be discussed in more detail here.

The proton motive force of Pma1 is used for the transport of ions and small molecules by ion channels and transporters, dissipating the very negative membrane potential and proton gradient. Futile cycles of ion transport have been studied in bacteria (e.g., Bond and Russel, 1998); these show that over a third of the energy could be dissipated by cycling ions through the cell membrane. For yeast, futile cycles of protons and anions under adaptation to weak organic acids explained about 30–40% reduction in intracellular ATP compared to growth inhibitory conditions (e.g., Ullah et al., 2013), while the percentage of dissipated energy needs more investigation. Futile cycles of ion transport for Na$^+$, K$^+$ and the other major ions are also well known for plant root cells, the estimates predict over a third and more of energy consumed by the cycles (Britto and Kronzucker, 2006; Malagoli et al., 2008 and later publications from the laboratory). At a first glance, futile cycles seem to be waste of energy and pitfalls of tight regulation. On the opposite, from the point of systems biology futile cycles are an additional mechanism of control and signaling and also a prerequisite for non-linear complex behavior of a biological system (Newsholme and Crabtree, 1976; Samoilov et al., 2005; Qian and Beard, 2006; Tolla et al., 2015). In yeast it is assumed that 20–60% of cell ATP is used by Pma1 (reviewed in Kahm, 2011). There are over 250 predicted transporters in the yeast genome (Paulsen et al., 1998) and almost 100 plasma membrane transporters of known function (Van Belle and André, 2001). Many of them transport small organic molecules, not ions, and can use the proton gradients and membrane potential created by Pma1. The "leak" current via membrane (non-selective cation channels and several other potential routes) and activity of transporters are regulated by membrane potential, calcium, ion concentrations etc. (Bihler et al., 1998; Roberts et al., 1999; reviewed in Ariño et al., 2010) and so are linked to the activity of Pma1 generating a feedback loop. There are still many questions as to which parameter or parameters are controlled and which mechanisms are involved in the regulation of ion transport. Voltage sensors and ligand or ion-binding sites of ion channels are the simplest structures that directly regulate ion currents depending on membrane potential, ion and ligand concentrations (e.g., Hille, 2001), making up regulatory networks. The study of various mutants is helpful in elucidating the interactions. Mutants with about 80% reduced expression of plasma membrane H$^+$-ATPase showed decreased rates of proton efflux and uptake of amino acids, while maintaining a stable membrane potential (measured by gradient of tetraphenylphosphonium) (Vallejo and Serrano, 1989). Mutants lacking outward potassium channel TOK1 had depolarized membrane potential (Maresova et al., 2006), while deletion of Trk1 and Trk2 potassium transporters hyperpolarized the membrane (Zahrádka and Sychrova, 2012).

**Ena1** is another ion pump in the yeast plasma membrane recognized as a putative $Na^+$ pump (Haro et al., 1991) removing $Na^+$ out of cytoplasm. Around 688 molecules per cell were estimated by protein expression analysis (Ghaemmaghami et al., 2003), which means that the potential maximal efflux is extremely low. Assuming pumping activity of Ena1 similar to Pma1 with 20–100 $H^+$/second (Serrano, 1988) or animal $Na^+/K^+$-ATPase with a turnover of 160 (Skou, 1998), an estimate of 100 $Na^+$ pumped out/second seems reasonable. Then the expected drop in intracellular $Na^+$ due to activity of Ena1 would be:

$$700 \text{ pumps/cell} * 100 \text{ } Na^+ \text{ions/(second*pump)}$$
$$= 70*10^3 Na^+ \text{ions/(cell*second) or about}$$
$$1 \text{ } \mu M/(cell*second)(\text{converting to concentrations using}$$
$$\text{Avogadro constant and cell volume) or 3.6 mM/(cell*h). (7)}$$

Thus, under control conditions Ena1 with a number of about 700 molecules/cell can hardly have an essential effect on sodium concentration. However, the transcription of Ena1 is activated by NaCl (Wieland et al., 1995) via multiple transduction pathways in dose-dependent manner (Márquez and Serrano, 1996). Microarray experiments revealed the complex pattern of activation (Posas et al., 2000 and http://transcriptome.ens.fr/ymgv/index.php for YDR040C), about 10-fold at most. Presumably the activation of transcription could be mirrored by increased number of protein molecules/cell. Interestingly, deletion of Ena1 in the yeast strain G19 resulted in dramatically retarded growth of yeast colonies on plates with added 0.5–1.3 M NaCl and even with added 1.8 M KCl especially at neutral pH 7.0 when compared with pH 3.6 and 5.5 (Bañuelos et al., 1998); therefore expression and interactions of Ena1 are expected to change more under long term salt stress.

## Ion Channels and Transporters

Major channels and transporters of yeast plasma membrane include: (1) potassium-selective outward rectifying ion channel TOK1, (2) non-selective cation channels (registered in electrophysiological experiments, but gene sequence(s) and protein structure(s) responsible for the current had not been identified so far), (3) potassium transporters Trk1 and Trk2, (4) sodium/potassium-proton antiporter Nha1, (5) sodiumphosphate symporter Pho89 (reviewed in Ariño et al., 2010), (6) voltage-gated calcium channel Cch1 (Fischer et al., 1997; Peiter et al., 2005; Teng et al., 2008, 2013) probably interacting with mechanosensitive cation-selective channel Mid1 (Kanzaki et al., 1999; Peiter et al., 2005), and (7) mechanosensitive ion channel with similar selectivity for cations and anions (Gustin et al., 1988). A brief description concerning the number of molecules per cell and estimated ion currents carried by them may help to understand their role in membrane transport properties. As a first approximation: sodium/potassium-proton antiporter Nha1 was visualized in experiments on protein expression in yeast using both GFP and TAP tags which estimated 1480 molecules/cell (about 15 molecules/$\mu m^2$), Mid1 was detected at 3210 molecules/cell (about 30 molecules/$\mu m^2$), while Trk1, Trk2, TOK1, Pho89, and Cch1 were not visualized or below the detection limit of 50 molecules/cell (Ghaemmaghami et al., 2003). It is worth mentioning that the number of specific protein molecules per cell and the protein abundance variation among cells in large populations of *S. cerevisiae* are partially determined genetically (Albert et al., 2014).

Though **TOK1** was not visualized in the study on protein expression (Ghaemmaghami et al., 2003), earlier electrophysiological experiments predicted about 50–100 active ion channels per a cell (Bertl and Slayman, 1992; Bertl et al., 1998). Potassium-selective outward rectifying ion channel TOK1 is the best characterized ion channel in yeast cells. First single channels recordings with yeast protoplasts revealed dominant conductance with high selectivity for potassium over sodium. The unitary conductance of about 13 pS (122 mM $K^+$ in pipette and outside medium) was inhibited by 20 mM $TEA^+$ and 10 mM $Mg^{2+}$ (Gustin et al., 1986). 17 pS potassium-selective and ATP-sensitive conductance was registered in $H^+$-ATPase mutants of *Saccharomyces cerevisiae* (150 KCl inside and outside configuration) (Ramirez et al., 1989). Several large (64 and 116 pS) potassium-selective conductances were revealed in yeast plasma membrane vesicles fused with planar bilayers. These conductances were inhibited by 10 mM $TEA^+$ and 10 mM $BaCl_2$ and presumably correspond to TOK1 (Gómez-Lagunas et al., 1989). Single channel recordings have been studied in detail (Bertl and Slayman, 1992; Bertl et al., 1993). Flickering behavior was found for the channel. The ion channel demonstrated high selectivity for potassium over sodium and was inhibited by increasing cytoplasmic calcium and by cytoplasmic sodium. A kinetic model with several closed and open states was proposed for gating (transition between open and closed states) of the channel (Bertl and Slayman, 1992; Bertl et al., 1993).

Whole cell recordings with yeast protoplasts provided more information about TOK1. Outward current via TOK1 was about 500 pA at +100 mV (175 mM pipette = cytoplasmic KCl and 150 mM external KCl) and demonstrated ATP-dependence (the current exhibited strong rundown within 60 min without ATP in pipette medium). There was no effect of external protons within pH 5.0–8.0, but inhibition by cytosolic acidification; inhibition by $TEA^+$ and $Ba^{2+}$ and gating by external potassium were also discovered (Bertl et al., 1998). Gating of the yeast potassium outward rectifier by external potassium was found for whole cell currents and confirmed for single channels. It was proposed (based on a previous model for the channel) that TOK1 has binding sites for potassium and plays a central role in osmoregulation and $K^+$-homeostasis in yeast (Bertl et al., 1998). Effects of TOK1 gating by external potassium and also amino acid residues responsible for the gating have been studied in detail using a heterologous expression system, oocytes of *Xenopus laevis* (Ketchum et al., 1995; Lesage et al., 1996; Loukin et al., 1997; Loukin and Saimi, 1999; Johansson and Blatt, 2006). It was an unexpected finding that TOK1 was slightly activated by volatile anaesthetic agents isoflurane, halothane and desflurane (Gray et al., 1998).

Simple estimates translate the current via the channel to ion concentrations: e.g., 10 pS correspond to 1.0 pA at 100 mV, the number of transported ions per second at 100 mV will be:

$$\approx 1{*}10^{-12} \text{coulombs/second}{*}1\text{second}{*}6.2{*}10^{18}$$
$$\text{monovalent ions/coulomb} \approx 6{*}10^{6} \text{ions} \quad (8)$$

Assuming a change in $K^+$ concentration from 100 to 40 mM within 10 min (600 s), we estimate that only one TOK1 molecule (with extremely low conductivity, at the level of patch clamp resolution) is sufficient to ensure the potassium ion fluxes.

Questions about the physiological role of TOK1 arose together with electrophysiogical characterization of the channel and were partially solved in experiments with yeast mutants. A yeast mutant with a deleted TOK1 gene had no potassium selective outward current (Bertl et al., 2003). Deletion of TOK1 depolarized cell membrane, while overexpression hyperpolarized it. These results were obtained in an assay using 3,3′dipropylthiacarbocyanine iodide fluorescence (Maresova et al., 2006). Potential participation of TOK1 in potassium uptake by yeast cells was demonstrated in yeast mutants with overexpressed TOK1 and deleted Trk1 and Trk2 genes (Fairman et al., 1999). Overexpression of TOK1 restored growth on plates with low (1 mM) potassium concentration and more than doubled potassium contents in cells cultured on 5.0 mM $K^+$ (Fairman et al., 1999). The results seem unusual at first glance: overexpression of outward rectifying ion channel has a positive effect on ion accumulation. A possible explanation for this is due to small inward current below E reversal for $K^+$, which was recorded in TOK1-overexpressing cells (Fairman et al., 1999). However, *tok1*1 mutants also had slightly higher potassium contents and did not differ in growth (cited according to Kahm, 2011). It is therefore proposed that TOK1 can contribute to controlling membrane potential around reversal potential for potassium (Fairman et al., 1999): the number of ions transferred via TOK1 [(Equation 8) for 100 ion channels but multiplied by smaller voltage and open probability for channels] is comparable to the number of ions pumped by Pma1 per second (Equation 6); so both Pma1 and TOK1 may play a role in altering the membrane potential.

**Non-selective cation current** in yeast cells is well studied by electrophysiology. Surprisingly high inward currents often exceeding 1 nA at −200 mV have been registered in yeast protoplasts (from larger tetraploid cells) when external divalent ions were reduced below 10 μM (Bihler et al., 1998, 2002). The currents demonstrated inward rectification increasing at more negative voltages and included time-dependent and timeindependent instantaneous components (Bihler et al., 1998, 2002). Slightly lower similar currents (about 300 pA at −140 mV) were recorded with 100 μM of external calcium and magnesium from spheroplasts of 4–5 μm in diameter (Roberts et al., 1999). The current was carried by monovalent cations and had similar selectivities for $K^+$ and $Rb^+$. Selectivity for $Na^+$ was about 50% less, while choline and $TEA^+$ were nearly impermeable cations (Roberts et al., 1999). Lowering intracellular potassium increased the magnitude of the current, so regulation by intracellular potassium was suggested (Roberts et al., 1999). While external pH within the range of 5.5–7.0 had no effect on the current (Roberts et al., 1999), pH 4.0 was found to be inhibitory (Bihler et al., 2002). Similar large inward currents have been recorded in another yeast, *Zygosaccharomyces bailii*; the conductance was not permeable to $TEA^+$ and was slightly inhibited by external 10 mM $Ca^{2+}$ or pH 4.0 (Demidchik et al., 2005). Nonselective cation channels are better studied in plants, where they have an essential role in salt tolerance and development. Cyclic nucleotide-gated ion channels and amino acid-gated channels make up the total cation non-selective currents in plants (Demidchik and Maathuis, 2007), while no candidate genes or proteins responsible for the current have been identified so far in yeast.

Obviously, regulation of yeast non-selective cation currents and changes in membrane potential will essentially change the influx of ions by the pathway, since 1 nA current carries per second:

$$1.0 \text{nA(measured at} -200 \text{ mV)corresponds to} \approx 1{*}10^{-9}$$
$$\text{coulombs/second}{*}1\text{second}{*}6.2{*}10^{18} \text{ions/coulomb}$$
$$= 6.2{*}10^{9} \text{ions}, \quad (9)$$

That is about 100% of cell potassium concentration per second for 6 μm yeast cells or 25% for 10 μm tetraploid cells. Probably under realistic physiological conditions the fluxes of ions via ion channels are influenced by kinetic factors and diffusion of ions to ion channels.

**Nha1** is an electroneutral or electrogenic cation/$H^+$ antiporter especially important for $Na^+$ efflux at low external pH values (Prior et al., 1996; Bañuelos et al., 1998; KinclovaZimmermannova et al., 2006). Assays with secretory vesicles evidence that during the transport cycle Nha1 transports with similar affinities one single ion of $Na^+$ or $K^+$ per more than one $H^+$ using energy of electrochemical gradient of $H^+$ (Ohgaki et al., 2005), but more experimental support is required. Though amino acids important for selectivity of Nha1 and protein domains essential for its regulation are known (Kinclová et al., 2001; Kinclova-Zimmermannova et al., 2005), mechanism and stoichiometry of ion transport by Nha1 still need to be elucidated. Potentially heterologous expression of Nha1 could be a step in the direction, however it was not successful so far (Volkov, personal communication).

Ion transporters have much lower transport rates compared to ion channels. Ion channels often transport over $10^6$ ions/(second*molecule), while transporters about 100–10,000 ions per second with the estimated theoretical limit of 100,000, due to the speed of protein conformational changes (e.g., Levitan and Kaczmarek, 2001, p. 72; Chakrapani and Auerbach, 2005). Similar to the Nha1 transporter is NhaA found in *E. coli*, which has a turnover number of about 1500 ions per second at pH 8.5 (Taglicht et al., 1991), so an estimate of 1000 transported ions/(second*transporter) in yeast seems reasonable to suggest.

Sodium transport out of cytoplasm by Nha1 would therefore be about:

$$1500 \text{Nha1 molecules/cell} * 1000 \text{ ions/(second*transporter)}$$
$$= 1.5*10^6 \text{ ions per second per a cell or (converting to concentrations using Avogadro constant and cell volume)}$$
$$\text{about } 25 \, \mu\text{M/(second*cell)} = 90 \text{ mM/(hour*cell)} \quad (10)$$

or from 9 to 900 mM/(hour*cell), adding potential 10-fold variation in estimated unknown yet possible transport rates by Nha1. This value is still not very high compared to sodium uptake via non-selective cation currents (see above), Ena1 also has low sodium efflux capacity (Equation 7). Indeed, the rapid increase of intracellular sodium to 150 mM within 100 min was observed in yeast cells subjected to 1 M NaCl stress (García et al., 1997).

Several possible scenarios of cell behavior under high sodium stress may include in that case:

(1) a shift of membrane potential to more positive values and consequently essential drop of the non-selective cation conductance, which will reduce sodium accumulation in cells below the toxic levels. However, increasing the membrane potential to more positive values will have a negative effect on potassium uptake via non-selective cation conductance and consequently on growth;
(2) to induce expression of Ena1 and Nha1 increasing sodium efflux from yeast cell;
(3) to use other channels and transporters (which have yet been not discovered) to expel sodium from the cells while maintaining a high potassium uptake rate.

Experiments with yeast mutants with deleted Nha1 demonstrate, however, a more complex pattern. Deletion of Nha1 had no effect on membrane potential assayed by the fluorescent dye 3,3′-dipropylthiacarbocyanine iodide and no effect on growth on plates in the presence of 200 mM NaCl (KinclovaZimmermannova et al., 2006). Higher concentrations of sodium chloride (0.5 M) and in particular potassium chloride (1.8 M) inhibited growth of a yeast strain without the *nha1* gene on plates at pH 3.6 (Bañuelos et al., 1998). Overexpression of Nha1 increased NaCl tolerance and decreased intracellular sodium concentration by 150 mM within 20 min in yeast cells preloaded with sodium. However, the number of Nha1 transporters per cell was not known for the overexpressing cells (Bañuelos et al., 1998). Overexpression of Nha1 also hyperpolarized the plasma membrane (Kinclova-Zimmermannova et al., 2006), potentially involving proton pump Pma1 since Nha1 presumably transports inside more $H^+$ per $Na^+$ or $K^+$ out (Ohgaki et al., 2005).

**Trk1 and Trk2** were isolated by a drop in potassium uptake by yeast cells from the external medium (Gaber et al., 1988; Ko and Gaber, 1991). Genes *trk1* and *trk2* are homologous to genes of HKT transporters, which are $K^+$ or $Na^+$ uniporters or $K^+/Na^+$ symporters (Mäser et al., 2002; Waters et al., 2013). However, some controversy appeared about their transport properties when both Trk1 and Trk2 potassium transporters were studied using *trk1*1 *trk2*1 mutants in patch clamp experiments. Deletion of *trk1* and *trk2* nearly abolished inward current at −200 mV, the transporters are therefore responsible for about −40 pA in whole cell configuration with 150 mM KCl + 10 $CaCl_2$ outside/175 mM KCl inside. Deletion of *trk1* alone reduced the inward current by 20 pA from −40 pA to −20 pA, while deletion of *trk2* abolished the current (Bertl et al., 2003). Surprisingly, in support of earlier electrophysiological experiments (Bihler et al., 1999) the current was initially attributed to $H^+$, not to $K^+$.

Simple estimates show that 40 pA per second will therefore be equivalent to:

$$40 \text{ pA at } -200 \text{ mV correspond to}$$
$$\approx 40*10^{-12} \text{ coulombs/second} * 1 \text{ second} * 6.2*10^{18}$$
$$\text{ions/coulomb} \approx 2.4*10^8 \text{ protons} \quad (11)$$

This calculation is nearly $10^5$ times higher than the estimated number of 6000 protons per yeast cell at pH 7 (Equation 4) and will need robust and fast buffering system to maintain pH homeostasis. Complex dynamics for proton transport from the liquid layers adjacent to the membrane (including diffusion, kinetics of transport etc.) will also be required, since at pH 7.5 (Trk1/Trk2 currents measured in Bertl et al., 2003) the 1000 times larger volume (100 nL, 10 diameters of the yeast cell) around the cell would include the number of $H^+$ equal to:

$$\approx 6.02*10^{23}/\text{mole} * 10^{-7} \text{L} * 3*10^{-8} \text{mole/L} \approx 18*10^8, \quad (12)$$

Only 7–8 times above the measured current per second. Fast transport of protons would need high-speed diffusion rates and a better understanding of the processes from the point of physical chemistry/biophysics of transport.

Further patch clamp studies with *S. cerevisiae* protoplasts revealed pH-dependent chloride conductance for both Trk1 and Trk2 rather than $H^+$ fluxes via the transporters (Kuroda et al., 2004; Rivetta et al., 2005, 2011). The inward ion current (Kuroda et al., 2004) did not depend on cation, but on chloride (or the other anion) concentrations in buffers for electrophysiological experiments. Small (below 5 pA) ion currents were also assumed to be associated with $K^+$-currents in protoplasts of *S. cerevisiae*. However, the current 5 pA at −200 mV would correspond per second to:

$$\approx 5*10^{-12} \text{ coulombs/second} * 1 \text{second} * 6.2*10^{18}$$

$$\text{ions/coulomb} \approx 3*10^7 \text{ ions.} \qquad (13)$$

This would require (1) 10,000 Trk1 and Trk2 transporters (assuming turnover being 1,000 ions/second per transporter) per cell while they have not been found in protein expression essays so far (e.g., Ghaemmaghami et al., 2003); (2) more experimental investigation since small anion currents still could not be completely excluded when measuring small whole cell currents (within the range of few pA).

Chloride conductance by Trk1-type transporters was shown for another yeast-like species, human pathogenic fungus *Candida albicans*, where it was strongly inhibited by the chloride currents blocker DIDS (4,4′-diisothiocyanatostilbene-2,2′-disulfonic acid) (Baev et al., 2004). Further experiments revealed that CaTrk1p transporter from *C. albicans* was homologous to Trk transporters from *S. cerevisiae* and CaTrk1p was responsible for low affinity $K^+$ uptake (indicated by $Rb^+$ uptake) in *C. albicans* (Baev et al., 2004; Miranda et al., 2009). However, the measured $K^+$ fluxes were estimated to be at the lower limit of resolution for patch clamp with protoplasts of *Candida albicans* (Miranda et al., 2009). Instead, large chloride conductance was measured in the protoplasts suggesting both chloride and potassium transport by CaTrk1p (Miranda et al., 2009).

The question remains as to whether Trk1 and Trk2 alone make up the proteins with a potassium-selective pore or if they are acting in cooperation with other proteins forming potassium uptake pathways (Ariño et al., 2010). Crystal structure of a similar type of transporter, TrkH from *Vibrio parahaemolyticus*, demonstrates a selectivity filter for $K^+$ and $Rb^+$ over $Na^+$ and $Li^+$ (Cao et al., 2011). Atomic scale and molecular dynamics modeling of Trk1 from *S. cerevisiae* confirmed homology with HKT transporters and revealed essential role of glycine residues within potassium selective filter region of the protein, sodium ions were inhibiting for $K^+$ transport (Zayats et al., 2015). However, still more experimental evidence is required and heterologous expression systems may also be useful for solving the puzzle. It is expected that the chloride conductance of Trk1 and Trk2 is masking the tiny potassium conductance which cannot therefore be easily measured electrophysiologically in yeast cells in whole cell configuration. Presumably chloride conductance is formed independently of $K^+$-conducting pathway and $K^+$ and $Cl^-$ fluxes are not influencing each other, so Trk proteins rather have dual transport functions than symport/antiport $K^+$ and $Cl^-$ ions.

Yeast mutants with deleted Trk1 and Trk2 transporters have hyperpolarized membrane potential assayed by fluorescent dyes 3,3′-dihexyloxacarbocyanin (Madrid et al., 1998) and 3,3′-dipropylthiacarbocyanide iodide, and also a more alkaline intracellular pH (by about 0.3 units) under high (50 mM) and low (15 μM) potassium in the liquid growth medium (Navarrete et al., 2010; Plášek et al., 2013). Proton efflux from the *trk1*1 *trk2*1 mutants was higher under 50 mM potassium conditions (Navarrete et al., 2010). It is interesting to mention that the effect of hyperpolarization was associated with $K^+$ and $Na^+$, but not with $Cl^-$ ions (according to results of their substitution by MES) (Madrid et al., 1998).

## Basic Feedbacks for Ion Transport Regulation in Yeast Cell

A simple sketch of participating elements, measured parameters and interactions between them for ion transport is given in **Figure 4**. A more detailed scheme will require complex models similar to metabolic models for yeast cell (e.g., Österlund et al., 2013 for metabolic control) with at least (1) determined and linking ion fluxes via distinct transport systems, (2) understanding and analysing non-linear interactions, and also (3) taking into account wide range of influencing physical factors (osmotic pressure, temperature, diffusion coefficients etc.) and chemical environment (pH, ion concentrations etc.) surrounding a yeast cell.

Effects of NaCl treatment under different external conditions with expected important pathways for ion transport are depicted at **Figure 5**. Among the present limitations are the knowledge gap between results obtained by different methods and extrapolation of experimental conclusions for higher ion concentrations. Electrophysiological patch clamp study provides evidence for fast kinetic changes; the data allow to reveal the role of individual types of ion channels or transporters, but yeast protoplasts are deprived of cell wall, do not have complete set of regulation feedbacks. Moreover, the patch clamp data are recorded under controlled conditions of experimental solutions, which traditionally do not use over 100–200 mM of $Na^+$, $K^+$ or other cations. Experiments with ion analysis of yeast cells or mutants in specific transport systems are with intact cells, the disadvantages are lower temporary resolution and net fluxes of definite ions; mutants may also have altered regulation of ion transport. It is reasonable to add that ion concentrations and pH in liquid layers adjacent to plasma membrane from the outside of cytoplasm are influenced by cell wall.

They include (1) changes of cell membrane potential by fixed electric charges in cell walls; (2) effects of cell walls on ion buffering and ion concentrations; (3) expected ion-rectifying properties of cell walls; (4) assumed effect of shrinking and swelling of cell walls on mechanosensitive ion channels. Since yeast cell walls are similar to cell walls of plants though less studied, several facts and phenomena about plant cell walls will be also given.

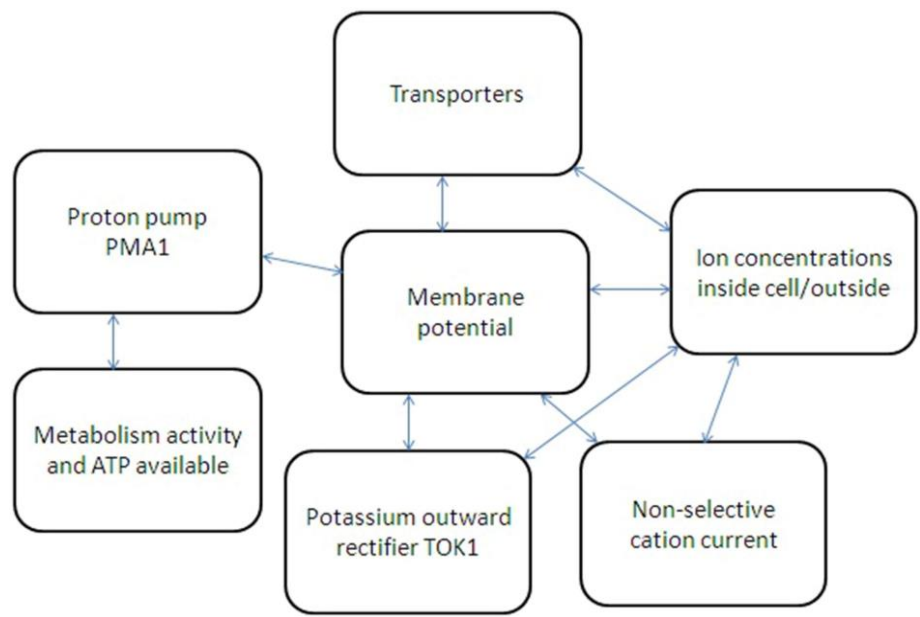

FIGURE 4 | **Simplified scheme of several feedbacks for ion transport regulation in yeast cell.** Metabolic activity and available ATP regulate proton pump Pma1, which influences membrane potential and also depends on membrane potential due to thermodynamic reasons. Non-selective cation currents and potassium current via potassium-selective outward rectifier TOK1 depend on cell membrane potential and ion concentration inside and outside yeast cell. Ion fluxes via transporters depend on ion concentrations; membrane potential is influencing them as well. The whole system of interacting factors and elements has non-linear feedbacks and links; therefore a system approach is required to describe the ion transport and its regulation.

## Cell Wall and Lipid Rafts Influence Ion Transport

The presence of a cell wall and the non-homogeneous structure of cell plasma membrane add more complexity for ion transport. Several as yet unexplored hypotheses should be considered based on the known physico-chemical properties of cell walls.

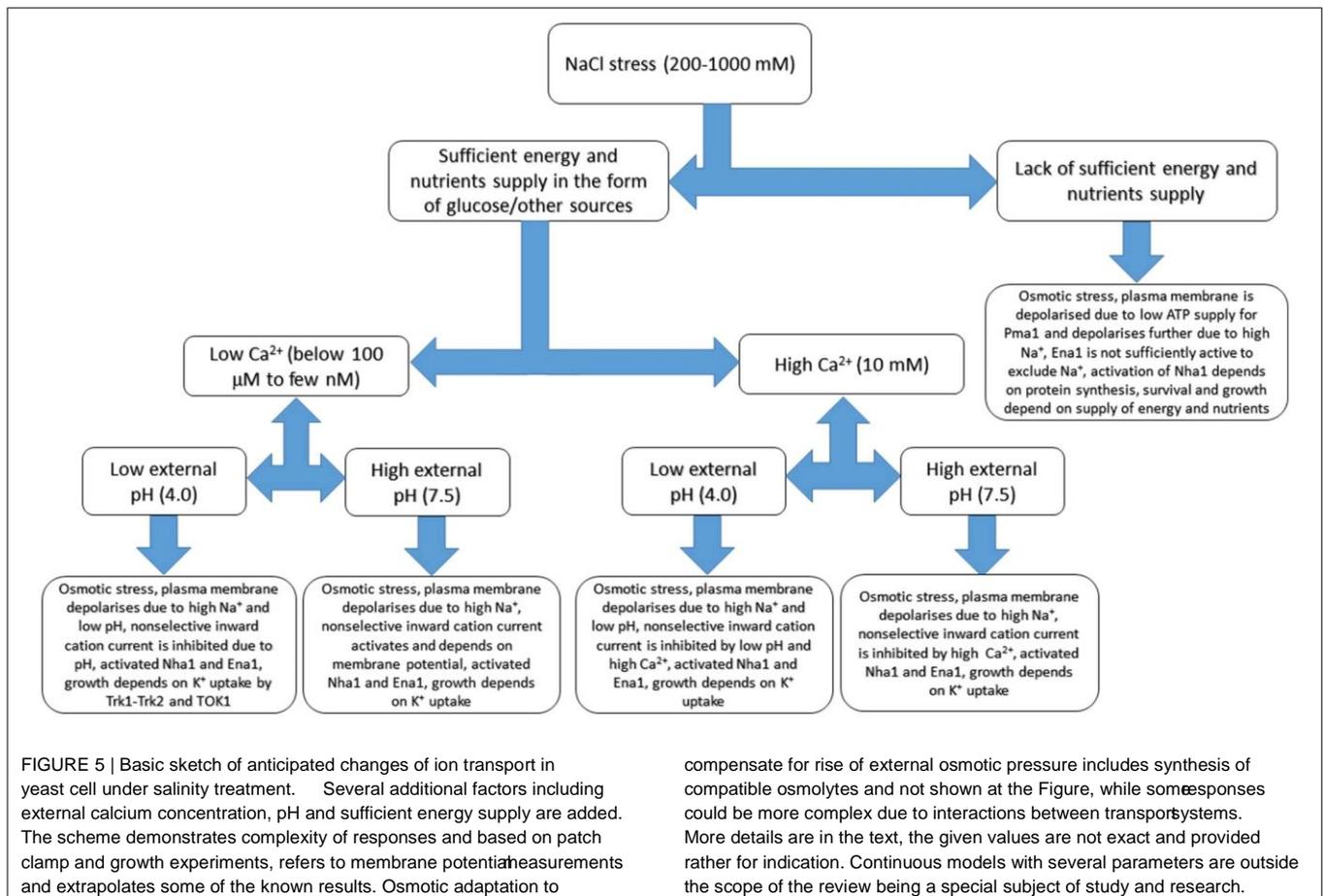

FIGURE 5 | Basic sketch of anticipated changes of ion transport in yeast cell under salinity treatment. Several additional factors including external calcium concentration, pH and sufficient energy supply are added. The scheme demonstrates complexity of responses and based on patch clamp and growth experiments, refers to membrane potential measurements and extrapolates some of the known results. Osmotic adaptation to compensate for rise of external osmotic pressure includes synthesis of compatible osmolytes and not shown at the Figure, while some responses could be more complex due to interactions between transport systems. More details are in the text, the given values are not exact and provided rather for indication. Continuous models with several parameters are outside the scope of the review being a special subject of study and research.

The yeast cell wall consists mainly of polysaccharides (reviewed in Cabib et al., 2001), has a thickness of about 0.1 μm (90–250 nm according to different sources, 9–25-fold thickness of plasma membrane) (Bowen et al., 1992; Smith et al., 2000; Stenson, 2009; Dupres et al., 2010) and keeps cell volume stable under sudden fluctuations of surrounding medium. Under usual conditions of low osmotic pressure of medium and high osmotic pressure of cell solution the cell wall is stretched and balances the turgor pressure (positive hydrostatic pressure of several bars (1 bar = 0.1 MPa) inside the cell to equilibrate chemical potentials of water caused by the difference of osmotic pressures).

The yeast cell wall consists of glucans, mannose polysaccharides, a few proteins and several per cents of chitin (Lipke and Ovalle, 1998; Cabib et al., 2001). The surface electrical charge of the cell is determined by polymers in the cell wall (chitin has charged amino groups, proteins have several charged chemical groups), and changes depending on the pH and ion composition of medium. Titration curves of surface charges of the yeast cell wall indicated wave around pH 7.1, value of $pK_{a2}$ for phosphate (Bowen et al., 1992). Much attention was then given to phosphomannans of the cell wall as they contain phosphate groups (Jayatissa and Rose, 1976).

Presumably the proton pump Pma1 and ion transporters have an effect on the charge distribution within the cell wall and visa versa ion transport is also influenced by the adjacent charged cell wall. The surface charge of a yeast cell can be measured by electrophoresis. For example, voltages 5–20 V of alternating current with frequencies in the range 0.03–5 Hz were applied to a suspension of yeast cells in distilled water. Cell of 5.7 μm in diameter was charged positively by about 10,000 elementary charges (personal communication of Ioan-Costin Stan in 2012: http://upcommons.upc.edu/pfc/bitstream/2099.1/10187/2/Master%20Thesis%20-%20Ioan.doc ), which equates to about 100 positive elementary charges/μm$^2$. Earlier results in buffered solutions, however, estimated a much higher surface charge (about 100 times larger in some cases) and demonstrated that electrophoretic mobility strongly depended on pH (negative charge at above pH 4), age of the culture and presence of phosphate in the growth medium (Eddy and Rudin, 1958). In these experiments isolated cell walls behaved similarly to whole yeast cells (Eddy and Rudin, 1958). In the other experiments electrophoretic mobility was also measured in buffered solutions; it dropped several times upon treatment with 60% hydrofluoric acid (HF) removing phosphate groups (Jayatissa and Rose, 1976).

The surface charge is expressed as the electric potential between the cell surface and the surrounding liquid medium. Zeta potential had been measured for yeast cells, showing values varying between 0 mV and −40 mV, being more negative in distilled water and at higher pH values (pH range 3–8 was studied). Usually zeta potential was not below −10 mV for cells in culture medium (Thonart et al., 1982; Bowen et al., 1992; Tálos et al., 2012).

Apart from slightly changing the measured membrane potential, the cell walls also have ion exchange properties, which can influence/buffer ion concentrations in the vicinity of plasma membrane and create local ion gradients. More is known for algae and plants, where cells possess similar cell walls with slightly different chemical composition. For example, in plants four types of ionogenic groups were found in cell walls isolated from lupine roots. Maximal cation-exchange capacity exceeded one mmole per gram of dry weight of cell walls (at pH over 10.8), while anion-exchange capacity was about 20 times lower (at pH below 2.8) (Meychik and Yermakov, 2001c). Ion exchange properties of plant cell walls from roots of lupine and horse beans essentially depended on pH and ion concentrations (Sentenac and Grignon, 1981). Algae had been studied earlier and demonstrated calcium exchange capacity about 40–400 mmole/(L cell wall) for charophytes (Tyree, 1968; reviewed in Hope and Walker, 1975) and cation exchange capacity over 2,5 mmole/g$^{-1}$ dry weight of cell walls for chlorophyte *Enteromorpha intestinalis* (Ritchie and Larkum, 1982). Cell walls of studied algae are thicker, averaging 0.3–1 μm (Wei and Lintilhac, 2007) and reported even up to 15 μm (Tyree, 1968).

Assuming similar range of cation-exchange capacity for yeast and then taking into account that cell wall makes up 15–30% of the dry weight of the cell and 25–50% of the volume based on calculations from electron micrographs (cited according to Lipke and Ovalle, 1998), it is simple to assess: about 1 mM per gram of dry weight is 1 M per kg of dry weight and about 200 mM per kg of fresh weight (1–5 is a reasonable fresh to dry ratio for yeast cell walls), similar to potassium concentration in yeast cell. Taking the volume of yeast cell wall to be about 20 fL (20% of 100 fL), the presumed maximal amount of cations bound by the cell wall would be:

$$20 \text{ fL} * 200 \text{ mM} = 4 \text{ fmole.} \qquad (14)$$

The suggested amount may be released depending on the pH of medium within the cell wall (therefore the proton pump Pma1 activity may be involved) and buffer ion transport fluxes.

It is worth mentioning again that in plants (roots of lupine, wheat and pea) cell walls shrink at higher ion concentrations and at lower pH values (for example, dry cell walls of lupin roots bound up to 10 times more water at pH 9 compared to pH 4 at 0.3 mM ionic strength and the same amount at 1 M ionic strength) (Meychik and Yermakov, 2001a,b). Moreover, cation-exchange capacity of plant roots (spinach) increased by about 30% during growth in high salt conditions (250 mM NaCl) (Meychik et al., 2006). High cation-exchange capacity of cell walls of plant roots was visualized in experiments using small organic cation methylene blue, it resulted in 100–700-fold higher concentration of methylene blue in cell walls than in ambient solution (Meychik et al., 2007). Furthermore, diffusion coefficients of methylene blue into cell walls differed between species (10 times higher in mungbean compared to wheat) and positively correlated with the number of carboxyl groups in cell wall structure and swelling of cell walls (amount of water absorbed by dried cell walls) (Meychik et al., 2007). The comparisons might be useful for studying and predicting the behavior of yeast cell walls.

A deeper knowledge of yeast cell walls and the variation between yeast strains and under different growth conditions may lead to a better understanding of high affinity cation (especially potassium) transport in yeast cells. Cell-wall mutants of yeast (e.g., described in Soltanian et al., 2007) could be useful in the study of this. Yeast cell walls are heterogeneous within a cell (Rösch et al., 2005) and may potentially have electrically rectifying properties. Considering yeast cell walls as charged porous medium one may suggest the essential influence on ion fluxes via plasma membrane (Lemaire et al., 2010). The complexity of cell walls with potential analogs to microand nanofluidics in electrically charged medium under applied voltages needs to be studied in more detail. Several descriptions and models had already been developed for plant and algal cell walls. These include Donnan free space with micro- and nanochannels and voltages down to -100 mV within it and also water free space, both constituting cell wall (e.g., Dainty and Hope, 1961; Hope and Walker, 1975; Beilby and Casanova, 2014).

The approach of considering cell wall and cell plasma membrane together as an interacting system may have implications on high affinity potassium transport in plants also. Inward rectifying potassium channels and transporters of HAK and HKT families are important for the pathway of transport in plants. For example, in *Arabidopsis thaliana* the potassium channel AKT1 is the high-affinity pathway for uptake of potassium from medium with concentrations as low as 10 μM while root cells had membrane potentials below −200 mV, sufficient for ensuring the electrochemical ion transport (Hirsch et al., 1998). Experimental evidence for effects of plant cell walls on calcium and proton fluxes was obtained for bean leaf mesophyll (Shabala and Newman, 2000). Isolated mesophyll protoplasts did not show NaCl-induced calcium efflux compared to mesophyll tissue; salt-induced $H^+$ efflux also differed between protoplasts compared to tissue. Presumably, all the $Ca^{2+}$ efflux over an hour of measurement was from calcium ion exchange stores of cell walls (Shabala and Newman, 2000). Similarly, transient $Ca^{2+}$ outward fluxes under certain pH changes were recorded for isolated cell walls of charophytes (Ryan et al., 1992). Further recent interesting observations have shown an essential step of iron binding to the cell wall in several algal species (though not in *S. cerevisiae* studied for comparison) before iron is taken up (Sutak et al., 2012).

Mechanical properties and turgor pressure of yeast cells have been measured by compression (Smith et al., 2000; Stenson,

2009; Stenson et al., 2011) and gives results around 100–180 MPa for mean Young elastic modulus. Turgor pressure in fission yeast was estimated at around 0.8 MPa (Minc et al., 2009). Elastic modulus of yeast cells looks higher at a first glance than usually measured for plant cells and their cell walls, the available studied similar system. For example, average volumetric elastic modulus of barley leaf cells was within 2–10 MPa and below 25 MPa, while turgor pressure was around 0.4–0.8 MPa (e.g., Volkov et al., 2007). However, methods of measurement are different for Young elastic modulus and volumetric elastic modulus even though both are expressed in units of pressure. Measurements of mechanical properties for oak leaf cells demonstrated a strong correlation between the two, with values of volumetric elastic modulus around 10 MPa corresponding to 100–200 MPa of
Young elastic modulus (Saito et al., 2006). More detailed results concerning turgor pressure and elastic modulus of different plant cells (where water relations and mechanical properties of cell walls are studied better so far) can be found in numerous reviews and publications (e.g., Hüsken et al., 1978; Steudle, 1993; Fricke and Peters, 2002). Turgor pressure in magnetotactic bacteria of the gram-negative species *Magnetospirillum gryphiswaldense* was estimated in the range of 0.09–0.15 MPa using atomic force microscopy (Arnoldi et al., 2000); similar or slightly higher turgor pressure of 0.08–0.5 MPa was measured for several other gramnegative bacteria (Beveridge, 1988; Walsby et al., 1995). Grampositive bacteria have 5–10 times higher turgor reaching 2–5 MPa (Beveridge, 1988; Doyle and Marquis, 1994), their cell walls are also much thicker being about 20–80 nm (e.g., Beveridge, 1988; Salton and Kim, 1996) compared to 5–10 nm in gram-negative bacteria (e.g., Salton and Kim, 1996).

Rapid changes in turgor pressure of yeast cells (influenced also by potential shrinking-swelling of cell wall) can activate mechanosensitive ion channels and induce ion flows via them, changing ion concentrations in different cell compartments. So far the well-studied yeast mechanosensitive ion channels include (1) mechanosensitive cation-selective channel Mid1 (Kanzaki et al., 1999; Peiter et al., 2005), (2) mechanosensitive ion channel with similar selectivity for cations and anions (Gustin et al., 1988), and also (3) the large (320 pS in 150 KCl in bath and 180 KCl in pipette solution) yeast vacuolar conductance resembling TRP-channels (Palmer et al., 2001; Zhou et al., 2003) with yet unknown functions (potentially the large trans-vacuolar currents under small changes in pressure could transport ions to cytoplasm for osmoregulation). Attempt to attribute activity of mechanosensitive ion channel with similar selectivity for cations and anions (Gustin et al., 1988) to Mid1 was not convincing (reviewed in Kung et al., 2010). The ion currents via mechanosensitive ion channels were registered in electrophysiological patch clamp experiments under slight positive (+) or negative (−) pressure applied to plasma or vacuolar membrane. The applied pressures were quite low compared to turgor pressure: −4 to −0.5 kPa for Mid1 (Kanzaki et al., 1999), +2.5 to +6.5 kPa and −8 to −0.5 kPa for mechanosensitive ion channel with similar selectivity for cations and anions (Gustin et al., 1988) and +0.6 to +6 kPa for mechanosensitive vacuolar Yvc1p channel (Zhou et al., 2003). Presumably even small alterations below 1% of turgor pressure could be amplified by ion fluxes via mechanosensitive ion channels, include calcium signaling (Kanzaki et al., 1999; Palmer et al., 2001; Zhou et al., 2003; Peiter et al., 2005) and further osmotic stress signaling to restore turgor pressure (e.g., reviewed in Hohmann, 2002; modeled in Muzzey et al., 2009; Schaber et al., 2010). Rapid transients of ion fluxes with resolution of seconds under osmotic stress and variation in responses between different cells are still to be measured for yeast cells with intact cell walls. Recent experiments using technique of microelectrode ion flux estimation (MIFE) revealed fast activation of $H^+$ efflux by over 3–4 times, change from $K^+$ efflux to influx and slight increase in $Mg^{2+}$ efflux under hyperosmotic stress (2 MPa by mannitol) in Gram-positive bacterium *Listeria monocytogenes* (Shabala et al., 2006), the MIFE technique is now being used for yeast cells (Ariño et al., 2014). The activity of mechanosensitive ion channels is also essentially determined by the rheology of membrane (Bavi et al., 2014) and so could be different in lipid raft compared to adjacent areas of plasma membrane.

**Lipid rafts** are found in yeast plasma membrane and the distribution of ion channels and transporters is not even within the membrane (e.g., reviewed in Malinsky et al., 2010). The Nha1 transporter was found to be associated with lipid rafts and requires sphingolipid for stable localization to the plasma membrane (Mitsui et al., 2009). The proton pump Pma1 was located to network structures, while proton transporters were in non-overlapping long-lived (over 10–30 min) 300 nm patches (Malínská et al., 2003; Grossmann et al., 2007), which changed upon membrane depolarization (Grossmann et al., 2007). More complex patchy pattern of plasma membrane had been discovered in yeast cells using total internal reflection fluorescence microscopy and following the location of 46 membrane proteins tagged with green fluorescent protein (GFP) (Spira et al., 2012). Another approach was to use atomic force microscopy (AFM) and the His-tagged transmembrane protein Wsc1. Scanning the cell surface with an AFM tip bearing $Ni^{2+}$nitriloacetate revealed clusters with a diameter of around 200 nm (Heinisch et al., 2010). The membrane protein Wsc1 was located within these clusters and stress conditions (deionized water or heat shock) increased the number of clusters and their size (Heinisch et al., 2010).

Lipid rafts with patchy localization of transporters can interact with the membrane skeleton, cell wall, regulatory proteins and cell cytoskeleton and make up local temporary ion gradients at the scale of several nanometers. The local non-equilibrium ion states depend on activity of transport systems and diffusion coefficients, it finally creates dynamic and live structure with fast fluctuations of ion currents and their complex regulation (**Figure 6**).

## Comparison of Yeast Cells and Large Excitable Cells

It is reasonable to ponder the difference between small yeast cells and excitable animal cells (**Figure 2**). The latter have a higher volume per unit of surface area (*vice versa* smaller surface per unit of volume), and use calcium and sodium ion channels with high conductance instead of immediate activity of electrogenic ion pumps for fast changes in membrane potential within hundreds of milliseconds and even much faster. Being more precise, the volume per unit of membrane electric capacitance has to be considered: the parameter is over 100 times higher for squid axon (Lecar et al., 1967) and about 5 times higher for cardiomyocytes (Satoh et al., 1996) compared to yeast cells. Moreover, the animal cells are functioning under controlled conditions of external medium (internal liquid of organisms) with buffered pH and relatively stable concentrations of $K^+$,

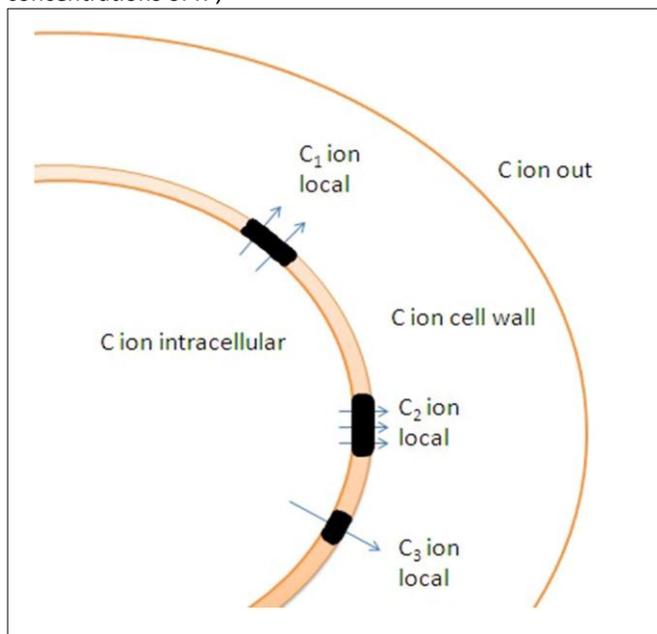

FIGURE 6 | Potential effects of cell wall and uneven distribution of ion channels and transporters in membrane on ion transport. Lipid rafts (shown in black color) are enriched with transporters. $C_1$, $C_2$, and $C_3$ indicate distinct local ion concentrations in the vicinity of lipid rafts caused by altered ion fluxes and ion buffering by cell wall. Specific rheological properties of lipid rafts result in additional differences in ion fluxes under changes in hydrostatic pressure due to non-identical activity of mechanosensitive ion channels within and outside the lipid rafts.

$Na^+$, $Ca^{2+}$, and $Cl^-$, but the limitations by cell geometry and by microscopic properties of ion pumps, transporters and ion channels are essential. Potentially to optimize ion fluxes the cells have choices of changing the number of membrane transport proteins and their regulation (e.g., yeast cells increase the density of plasma membrane proton pump Pma1 up to higher limits) or choosing specific transport proteins to fulfill the required physiological/biochemical demands. High density of sodium channels with high conductance is used in specialized segments of neurons (e.g., Kole et al., 2008). Imagining similar fluxes for a yeast cell with sodium channel density 2,500 pS $\mu m^{-2}$, it is simple to recalculate the theoretical sodium flux under the conditions of membrane potential equal −100 mV:

2500 pS $\mu m^{-2}$ to 100 $\mu m^2$ of yeast cell is 250,000 pS/cell, under −100 mV it would correspond to −25 nA/cell. Taking into account that the number of monovalent cations per Ampere*second = coulomb is equal to $6.2*10^{18}$ ions/coulomb, the ion current per second is equivalent to $\approx 1.6*10^{11}$ $Na^+$ ions. Note that it is $Na^+$ current, but it is more than 25 times over the total number of, e.g., $K^+$ ions in a yeast cell or 40 ms of the ion current are sufficient to carry the same number of cations which are present in the cell. Yeast cells do not have the sort of ion channels with high conductance and density, otherwise they will not be able to keep ion homeostasis under conditions of undetermined and changing ion composition of surrounding medium. Instead yeast cells use plasma membrane pump to keep required membrane potential and ion transporters with much lower transport rate to support smaller ion fluxes.

Another comparison is also interesting. Large cells of charophyte algae (**Figure 2**) have active plasma membrane $H^+$ATPase and exhibit action potentials (reviewed in Beilby, 2007; Beilby and Casanova, 2014). The group of multicellular algae with cells up to 10 cm long and 2 mm in diameter (Johnson et al., 2002) demonstrates slow action potentials within tens of seconds based on changes in chloride and calcium electric conductances (Beilby, 2007). Similar to yeast cells the large charophytes use $H^+$-ATPase for shifting membrane potential to more negative values; charophytes have membrane potential down to −350 mV (comprehensively described in the book: Beilby and Casanova, 2014). Charophyte cells have several physiological "states" of plasma membrane depending on activity of ion transport systems. "Pump state" corresponds to active $H^+$-ATPase and membrane potential below -200 mV, "$K^+$ state" is dominated by $K^+$ conductances, while "$H^+/OH^-$ state" is determined by passive $H^+/OH^-$ conductances (Beilby and Casanova, 2014). It is questionable whether yeast cells may also have several physiological states of membrane. Slow compared to animal cells action potentials in charophytes are probably explained by their evolution and set of ion channels required for variable unstable ion environment despite the presence of shielding and ion buffering by thick cell walls; the situation needs further modeling and investigation.

The reasons for the difference in ion transport systems are of dual nature: (1) pure biophysical constrains determine the ion transport fluxes and also (2) the history of species and evolution. A few more comparisons with yeast cells are useful. Human erythrocytes have cell volume similar to yeast cells, possess an electrogenic $Na^+/K^+$-pump, have low membrane potential (about −10 mV when measured by glass electrode) (Jay and Burton, 1969) and several ion channels, but do not use the channels under usual conditions, instead relying on ion pump and transporters to keep ion homeostasis (reviewed in Thomas et al., 2011). Changing the membrane potential from −10 to −90 mV by sodium ionophore has no effect on swelling-activated potassium transport in erythrocytes (Kaji, 1993). Another example of small animals cells are spermatozoa. The cells have volume about 20–30 fL, their measured membrane potential is usually depolarized to around −40 mV and was reported from −80 to +13 mV depending on method of measurement and conditions (Rink, 1977; Guzmán-Grenfell et al., 2000; reviewed in Darszon et al., 2006; Navarro et al., 2007). The membrane potential of spermatozoa is easily shifted to more negative voltages by activated electrogenic $Na^+/K^+$pump (Guzmán-Grenfell et al., 2000). Moreover, $Na^+/K^+$-pump is essential for lower membrane potential and also for fertility (Jimenez et al., 2011), though the cells are equipped with a range of ion channels and additionally depend on potassium channel to regulate the membrane potential (Darszon et al., 2006; Navarro et al., 2007).

Evolutionary aspects of differences in ion transport are quite complicated and could not be explained easily. Excitable cells usually have sodium channels, which have evolved from calcium channels and predated the origin of the nervous system in animals (Liebeskind et al., 2011). Interestingly, action potentials have been found in some marine algae, e.g., sodium/calciumbased action potential in *Odontella sinensis* (Taylor, 2009), which has a volume about 300 times larger than yeast. Chloride/calcium based action potentials in large charophytes are studied better (Beilby, 2007). Sodium channels and action potentials are not known in fungi so far, though sodium channels have surprisingly been found in the bacterium *Bacillus halodurans* (Ren et al., 2001). The bacterium, which is about 200 times smaller than the yeast cell by volume, can grow in a saline environment (2M NaCl) (Takami and Horikoshi, 1999). Opening of just one channel with 12 pS conductance (measured in 140 mM NaCl using expression in CHO-K1 cells) would carry large $Na^+$ currents and may essentially change sodium concentration in cytoplasm of the bacterium. The biological function of the channel is not known, but it may be related to driving the flagellar motor (Ren et al., 2001).

Biophysical limitations are easier for understanding, modeling and for an attempt of synthetic engineering (**Figure 2**), however, the more detailed description is outside the scope of the text.

## Conclusions

The yeast plasma membrane is different from the plasma membrane of excitable animal cells: yeast cells have no abundant sodium and calcium channels with high conductance. The membrane potential of yeast cells can be easily controlled by copious plasma membrane ATPase Pma1, potentially it can change the membrane potential of the small cells by several hundred millivolts within a few seconds. Transporters are also involved in the regulation of membrane potential. Non-selective cation current and outward rectifying potassium selective channels may serve as safety valves against the sharp changes in membrane potential. The small volume of yeast cells, their relatively large surface/volume ratio and high transport capacity of Pma1 make the proton pump, not ion channels, the main element of ion transport. Therefore, the regulation of proton pump and ion transporters is vital for ion homeostasis in yeast cells. The basic principles of ion transport in yeast cells may be applicable to small cells and organelles. Rapid processes of ion transport, within scale of seconds, can essentially change ion concentrations in the cells taking into account their small volume. The longterm scale of minutes and hours needs other mechanisms and sensors for regulation and homeostasis. A high cationexchange capacity of cells walls, location of ion channels and transporters within lipid rafts and changes in turgor pressure may essentially influence ion transport fluxes and make the situation more complex and yet unpredictable at smaller spatial scales.

## Acknowledgments

The Author expresses sincere thanks to his colleagues Gareth Williams, Christopher Palmer, Leonid Krivitsky, Mary Beilby, and Louise Krska, Reviewers and the Editors for careful reading and comments on the manuscript and apologizes for not citing all the relevant literature sources due to limitations by the text frames of the Review.

## References

Albert, A., Yenush, L., Gil-Mascarell, M. R., Rodriguez, P. L., Patel, S., MartínezRipoll, M., et al. (2000). X-ray structure of yeast Hal2p, a major target of lithium and sodium toxicity, and identification of framework interactions determining cation sensitivity. *J. Mol. Biol.* 295, 927–938. doi: 10.1006/jmbi.1999.3408

Albert, F. W., Treusch, S., Shockley, A. H., Bloom, J. S., and Kruglyak, L. (2014). Genetics of single-cell protein abundance variation in large yeast populations. *Nature* 506, 494–497. doi: 10.1038/nature12904

Albrecht-Buehler, G. (1990). In defense of "nonmolecular" cell biology. *Int. Rev. Cytol.* 120, 191–241.

Alemán, F., Caballero, F., Ródenas, R., Rivero, R. M., Martínez, V., and Rubio, F. (2014). The F130S point mutation in the Arabidopsis high-affinity $K^+$ transporter AtHAK5 increases $K^+$ over $Na^+$ and $Cs^+$ selectivity and confers $Na^+$ and $Cs^+$ tolerance to yeast under heterologous expression. *Front. Plant Sci.* 5:430. doi: 10.3389/fpls.2014.00430

Ambesi, A., Miranda, M., Petrov, V. V., and Slayman, C. W. (2000). Biogenesis and function of the yeast plasma membrane $H^+$-ATPase. *J. Exp. Biol.* 203, 155–160. Available online at: http://jeb.biologists.org/content/203/1/155.long

Ariño, J., Aydar, E., Drulhe, S., Ganser, D., Jorrín, J., Kahm, M., et al. (2014). Systems biology of monovalent cation homeostasis in yeast: the Translucent contribution. *Adv. Microb. Physiol.* 64, 1–63. doi: 10.1016/B978-0-12-8001431.00001-4

Ariño, J., Ramos, J., and Sychrová, H. (2010). Alkali metal cation transport and homeostasis in yeasts. *Microbiol. Mol. Biol. Rev.* 74, 95–120. doi: 10.1128/MMBR.00042-09

Arnoldi, M., Fritz, M., Bäuerlein, E., Radmacher, M., Sackmann, E., and Boulbitch, A. (2000). Bacterial turgor pressure can be measured by atomic force microscopy. *Phys. Rev. E* 62, 1034–1044. doi: 10.1103/PhysRevE.62.1034

Ashe, M. P., De Long, S. K., and Sachs, A. B. (2000). Glucose depletion rapidly inhibits translation initiation in yeast. *Mol. Biol. Cell.* 11, 833–848. doi: 10.1091/mbc.11.3.833

Auer, M., Scarborough, G. A., and Kühlbrandt, W. (1998). Three-dimensional map of the plasma membrane H[+]-ATPase in the open conformation. *Nature* 392, 840–843. doi: 10.1038/33967

Babu, M., Vlasblom, J., Pu, S., Guo, X., Graham, C., Bean, B. D., et al. (2012). Interaction landscape of membrane-protein complexes in *Saccharomyces cerevisiae*. *Nature* 489, 585–589. doi: 10.1038/nature11354

Baev, D., Rivetta, A., Vylkova, S., Sun, J. N., Zeng, G. F., Slayman, C. L., et al. (2004). The TRK1 potassium transporter is the critical effector for killing of *Candida albicans* by the cationic protein, Histatin 5. *J. Biol. Chem.* 279, 55060–55072. doi: 10.1074/jbc.M411031200

Bakker, R., Dobbelmann, J., and Borst-Pauwels, G. W. F. H. (1986). Membrane potential in the yeast *Endomyces magnusii* measured by microelectrodes and TPP[+] distribution. *Biochim. Biophys. Acta* 861, 205–209. doi: 10.1016/00052736(86)90421-9

Bañuelos, M. A., Sychrová, H., Bleykasten-Grosshans, C., Souciet, J. L., and Potier, S. (1998). The Nha1 antiporter of *Saccharomyces cerevisiae* mediates sodium and potassium efflux. *Microbiology* 144, 2749–2758.

Bavi, N., Nakayama, Y., Bavi, O., Cox, C. D., Qin, Q. H., and Martinac, B. (2014). Biophysical implications of lipid bilayer rheometry for mechanosensitive channels. *Proc. Natl. Acad. Sci. U.S.A.* 111, 13864–13869. doi: 10.1073/pnas.1409011111

Beilby, M. J. (2007). Action potential in charophytes. *Int. Rev. Cytol.* 257, 43–82. doi: 10.1016/s0074-7696(07)57002-6

Beilby, M. J., and Casanova, M. T. (2014). *The Physiology of Characean Cells*. Heidelberg; New York; Dordrecht; London: Springer, 205.

Bertl, A., Bihler, H., Reid, J. D., Kettner, C., and Slayman, C. L. (1998). Physiological characterization of the yeast plasma membrane outward rectifying K+ channel, DUK1 (TOK1), in situ. *J. Membrane Biol.* 162, 67–80. doi: 10.1007/s002329900343

Bertl, A., Ramos, J., Ludwig, J., Lichtenberg-Fraté, H., Reid, J., Bihler, H., et al. (2003). Characterization of potassium transport in wild-type and isogenic yeast strains carrying all combinations of *trk1*, *trk2* and *tok1* null mutations. *Mol. Microbiol.* 47, 767–780. doi: 10.1046/j.1365-2958.2003.03335.x

Bertl, A., and Slayman, C. L. (1992). Complex modulation of cation channels in the tonoplast and plasma membrane of *Saccharomyces cerevisiae*: single-channel studies. *J. Exp. Biol.* 172, 271–287.

Bertl, A., Slayman, C. L., and Gradmann, D. (1993). Gating and conductance in an outward-rectifying K[+] channel from the plasma membrane of *Saccharomyces cerevisiae*. *J. Membr. Biol.* 132, 183–199. doi: 10.1007/BF00235737

Beveridge, T. J. (1988). The bacterial surface: general considerations towards design and function. *Can. J. Microbiol.* 34, 363–372. doi: 10.1139/m88-067

Bihler, H., Gaber, R. F., Slayman, C. L., and Bertl, A. (1999). The presumed potassium carrier Trk2p in *Saccharomyces cerevisiae* determines an H[+]-dependent, K[+]-independent current. *FEBS Lett.* 447, 115–120. doi: 10.1016/S0014-5793(99)00281-1

Bihler, H., Slayman, C. L., and Bertl, A. (1998). NSC1: a novel high-current inward rectifier for cations in the plasma membrane of *Saccharomyces cerevisiae*. *FEBS Lett.* 432, 59–64. doi: 10.1016/S0014-5793(98)00832-1

Bihler, H., Slayman, C. L., and Bertl, A. (2002). Low-affinity potassium uptake by *Saccharomyces cerevisiae* is mediated by NSC1, a calcium-blocked non-specific cation channel. *Biochim. Biophys. Acta* 1558, 109–118. doi: 10.1016/S00052736(01)00414-X

Blatt, M. R., and Slayman, C. L. (1983). KCl leakage from microelectrodes and its impact on the membrane parameters of a nonexcitable cell. *J. Membrane Biol.* 72, 223–234. doi: 10.1007/BF01870589

Blatt, M. R., and Slayman, C. L. (1987). Role of "active" potassium transport in the regulation of cytoplasmic pH by nonanimal cells. *Proc. Natl. Acad. Sci. U.S.A.* 84, 2737–2741. doi: 10.1073/pnas.84.9.2737

Bond, D. R., and Russel, J. B. (1998). Relationship between Intracellular Phosphate, Proton Motive Force, and Rate of Nongrowth Energy Dissipation (Energy Spilling) in *Streptococcus bovis* JB1. *Appl. Environ. Microbiol.* 64, 976–981.

Borodina, I., and Nielsen, J. (2014). Advances in metabolic engineering of yeast *Saccharomyces cerevisiae* for production of chemicals. *Biotechnol. J.* 9, 609–620. doi: 10.1002/biot.201300445

Borst-Pauwels, G. W. (1981). Ion transport in yeast. *Biochim. Biophys. Acta* 650, 88–127. doi: 10.1016/0304-4157(81)90002-2

Bowen, W. R., Sabuni, H. A., and Ventham, T. J. (1992). Studies of the cell-wall properties of *Saccharomyces cerevisiae* during fermentation. *Biotechnol. Bioeng.* 40, 1309–1318. doi: 10.1002/bit.260401104

Brindle, K., Braddock, P., and Fulton, S. (1990). [31]P NMR measurements of the ADP concentration in yeast cells genetically modified to express creatine kinase. *Biochemistry* 29, 3295–3302. doi: 10.1021/bi00465a021

Britto, D. T., and Kronzucker, H. J. (2006). Futile cycling at the plasma membrane: a hallmark of low-affinity nutrient transport. *Trends Plant Sci.* 11, 529–534. doi: 10.1016/j.tplants.2006.09.011

Broach, J. R. (2012). Nutritional control of growth and development in yeast. *Genetics* 192, 73–105. doi: 10.1534/genetics.111.135731

Brückner, A., Polge, C., Lentze, N., Auerbach, D., and Schlattner, U. (2009). Yeast two-hybrid, a powerful tool for systems biology. *Int. J. Mol. Sci.* 10, 2763–2788. doi: 10.3390/ijms10062763

Cabib, E., Roh, D. H., Schmidt, M., Crotti, L. B., and Varma, A. (2001). The yeast cell wall and septum as paradigms of cell growth and morphogenesis. *J. Biol. Chem.* 276, 19679–19682. doi: 10.1074/jbc.R000031200

Cao, Y., Jin, X., Huang, H., Derebe, M. G., Levin, E. J., Kabaleeswaran, V., et al. (2011). Crystal structure of a potassium ion transporter, TrkH. *Nature* 471, 336–340. doi: 10.1038/nature09731

Chakrapani, S., and Auerbach, A. (2005). A speed limit for conformational change of an allosteric membrane protein. *Proc. Natl. Acad. Sci. U.S.A.* 102, 87–92. doi: 10.1073/pnas.0406777102

Cui, J., and Kaandorp, J. A. (2006). Mathematical modeling of calcium homeostasis in yeast cells. *Cell Calcium.* 39, 337–348. doi: 10.1016/j.ceca.2005.12.001

Curry, M. R., Millar, J. D., Tamuli, S. M., and Watson, P. F. (1996). Surface area and volume measurements for ram and human spermatozoa. *Biol. Reprod.* 55, 1325–1332. doi: 10.1095/biolreprod55.6.1325

Dainty, J., and Hope, A. B. (1961). The electric double layer and the donnan equilibrium in relation to plant cell walls. *Aust. J. Biol. Sci.* 14, 541–551.

Darszon, A., Acevedo, J. J., Galindo, B. E., Hernández-González, E. O., Nishigaki, T., Treviño, C. L., et al. (2006). Sperm channel diversity and functional multiplicity. *Reproduction* 131, 977–988. doi: 10.1530/rep.1.00612

Demidchik, V., and Maathuis, F. J. (2007). Physiological roles of nonselective cation channels in plants: from salt stress to signalling and development. *New Phytol.* 175, 87–404. doi: 10.1111/j.1469-8137.2007.02128.x

Demidchik, V., Macpherson, N., and Davies, J. M. (2005). Potassium transport at the plasma membrane of the food spoilage yeast *Zygosaccharomyces bailii*. *Yeast* 22, 21–29. doi: 10.1002/yea.1194

Doyle, R. J., and Marquis, R. E. (1994). Elastic, flexible peptidoglycan and bacterial cell wall properties. *Trends Microbiol.* 2, 57–60. doi: 10.1016/0966842X(94)90127-9

Dupres, V., Dufrêne, Y. F., and Heinisch, J. J. (2010). Measuring cell wall thickness in living yeast cells using single molecular rulers. *ACS Nano* 4, 5498–5504. doi: 10.1021/nn101598v

Eddy, A. A., and Rudin, A. D. (1958). The structure of the yeast cell wall. I. Identification of charged groups at the surface. *Proc. R. Soc. Lond. Ser. B Biol. Sci.* 148, 419–432. doi: 10.1098/rspb.1958.0035

Eraso, P., Cid, A., and Serrano, R. (1987). Tight control of the amount of yeast plasma membrane ATPase during changes in growth conditions and gene dosage. *FEBS Lett.* 224, 193–197. doi: 10.1016/0014-5793(87)80446-5

Fairman, C., Zhou, X., and Kung, C. (1999). Potassium uptake through the TOK1 K[+] channel in the budding yeast. *J. Membr. Biol.* 168, 149–157. doi: 10.1007/s002329900505

Felle, H., Porter, J. S., Slayman, C. L., and Kaback, H. R. (1980). Quantitative measurements of membrane potential in *Escherichia coli*. *Biochemistry* 19, 3585–3590. doi: 10.1021/bi00556a026

Fields, S., and Song, O. (1989). A novel genetic system to detect protein-protein interactions. *Nature* 340, 245–246. doi: 10.1038/340245a0

Fischer, M., Schnell, N., Chattaway, J., Davies, P., Dixon, G., and Sanders, D. (1997). The *Saccharomyces cerevisiae* CCH1 gene is involved in calcium influx and mating. *FEBS Lett.* 419, 259–262. doi: 10.1016/S0014-5793(97)01466-X

Fricke, W., and Peters, W. S. (2002). The biophysics of leaf growth in saltstressed barley. A study at the cell level. *Plant Physiol.* 129, 374–388. doi: 10.1104/pp.001164

Gaber, R. F., Styles, C. A., and Fink, G. R. (1988). TRK1 encodes a plasma membrane protein required for high-affinity potassium transport in *Saccharomyces cerevisiae*. *Mol. Cell. Biol.* 8, 2848–2859.

García, M. J., Ríos, G., Ali, R., Bellés, J. M., and Serrano, R. (1997). Comparative physiology of salt tolerance in *Candida tropicalis* and *Saccharomyces cerevisiae*. *Microbiology* 143, 1125–1131. doi: 10.1099/00221287-143-4-1125

Ghaemmaghami, S., Huh, W. K., Bower, K., Howson, R. W., Belle, A., Dephoure, N., et al. (2003). Global analysis of protein expression in yeast. *Nature* 425, 737–741. doi: 10.1038/nature02046

Gómez-Lagunas, F., Peña, A., Liévano, A., and Darszon, A. (1989). Incorporation of ionic channels from yeast plasma membranes into black lipid membranes. *Biophys. J.* 56, 115–119.

Goossens, A., de La Fuente, N., Forment, J., Serrano, R., and Portillo, F. (2000). Regulation of yeast $H^+$-ATPase by protein kinases belonging to a family dedicated to activation of plasma membrane transporters. *Mol. Cell. Biol.* 20, 7654–7661. doi: 10.1128/MCB.20.20.7654-7661.2000

Gout, E., Rébeillé, F., Douce, R., and Bligny, R. (2014). Interplay of $Mg^{2+}$, ADP, and ATP in the cytosol and mitochondria: unravelling the role of $Mg^{2+}$ in cell respiration. *Proc. Natl. Acad. Sci. U.S.A.* 111, E4560–E4567. doi: 10.1073/pnas.1406251111

Gray, A. T., Winegar, B. D., Leonoudakis, D. J., Forsayeth, J. R., and Yost, C. S. (1998). TOK1 is a volatile anesthetic stimulated $K^+$ channel. *Anesthesiology* 88, 1076–1084. doi: 10.1097/00000542-199804000-00029

Grossmann, G., Opekarova, M., Malinsky, J., Weig-Meckl, I., and Tanner, W. (2007). Membrane potential governs lateral segregation of plasma membrane proteins and lipids in yeast. *EMBO J.* 26, 1–8. doi: 10.1038/sj.emboj.7601466

Gustin, M. C., Martinac, B., Saimi, Y., Culbertson, M. R., and Kung, C. (1986). Ion channels in yeast. *Science* 233, 1195–1197. doi: 10.1126/science.2426783

Gustin, M. C., Zhou, X. L., Martinac, B., and Kung, C. (1988). A mechanosensitive ion channel in the yeast plasma membrane. *Science* 242, 762–765. doi: 10.1126/science.2460920

Guzmán-Grenfell, A. M., Bonilla-Hernández, M. A., and González-Martínez, M. T. (2000). Glucose induces a $Na^+,K^+$-ATPase-dependent transient hyperpolarization in human sperm. I. Induction of changes in plasma membrane potential by the proton ionophore CCCP. *Biochim. Biophys. Acta* 1464, 188–198. doi: 10.1016/S0005-2736(99)00247-3

Haro, R., Garciadeblas, B., and Rodríguez-Navarro, A. (1991). A novel P-type ATPase from yeast involved in sodium transport. *FEBS Lett.* 291, 189–191. doi: 10.1016/0014-5793(91)81280-L

Hasenbrink, G., Kolacna, L., Ludwig, J., Sychrova, H., Kschischo, M., and Lichtenberg-Fraté, H. (2007). Ring test assessment of the mKir2.1 growth based assay in *Saccharomyces cerevisiae* using parametric models and modelfree fits. *Appl. Microbiol. Biotechnol.* 73, 1212–1221. doi: 10.1007/s00253-00 6-0589-x

Hasenbrink, G., Schwarzer, S., Kolacna, L., Ludwig, J., Sychrova, H., and Lichtenberg-Fraté, H. (2005). Analysis of the mKir2.1 channel activity in potassium influx defective *Saccharomyces cerevisiae* strains determined as changes in growth characteristics. *FEBS Lett.* 579, 1723–1731. doi: 10.1016/j.febslet.2005.02.025

Hasenbrink, G., Sievernich, A., Wildt, L., Ludwig, J., and Lichtenberg-Fraté, H. (2006). Estrogenic effects of natural and synthetic compounds including tibolone assessed in *Saccharomyces cerevisiae* expressing the human estrogen α and β receptors. *FASEB J.* 20, 1552–1554. doi: 10.1096/fj.05-5413fje

Heinisch, J. J., Dupres, V., Wilk, S., Jendretzki, A., and Dufrêne, Y. F. (2010). Single-molecule atomic force microscopy reveals clustering of the yeast plasma-membrane sensor Wsc1. *PLoS ONE* 5:e11104. doi: 10.1371/journal.pone.0011104

Hille, B. (2001). *Ion Channels and Excitable Membranes, 3rd Edn*. Sunderland, MA: Sinauer Associates, 814.

Hirsch, R. E., Lewis, B. D., Spalding, E. P., and Sussman, M. R. (1998). A role for the AKT1 potassium channel in plant nutrition. *Science* 280, 918–921. doi: 10.1126/science.280.5365.918

Hodgkin, A. L., and Huxley, A. F. (1952). A quantitative description of membrane current and its application to conduction and excitation in nerve. *J. Physiol.* 117, 500–544. doi: 10.1113/jphysiol.1952.sp004764

Höfer, M., and Novacky, A. (1986). Measurement of plasma membrane potentials of yeast cells with glass microelectrodes. *Biochim. Biophys. Acta* 862, 372–378. doi: 10.1016/0005-2736(86)90240-3

Hohmann, S. (2002). Osmotic stress signaling and osmoadaptation in yeasts. *Microbiol. Mol. Biol. Rev.* 66, 300–372. doi: 10.1128/MMBR.66.2.300-372.2002

Hope, A. B., and Walker, N. A. (1975). *The Physiology of Giant Algal Cells*. Cambridge: Cambridge University Press, 226.

Hüsken, D., Steudle, E., and Zimmermann, U. (1978). Pressure probe technique for measuring water relations of cells in higher plants. *Plant Physiol.* 61, 158–163. doi: 10.1104/pp.61.2.158

Jay, A. W. (1975). Geometry of the human erythrocyte. I. Effect of albumin on cell geometry. *Biophys. J.* 15, 205–222. doi: 10.1016/S0006-3495(75)85812-7

Jay, A. W. L., and Burton, A. C. (1969). Direct measurement of potential difference across the human red blood cell membrane. *Biophys. J.* 9, 115–121. doi: 10.1016/S0006-3495(69)86372-1

Jayatissa, P. M., and Rose, A. H. (1976). Role of wall phosphomannan in flocculation of *Saccharomyces cerevisiae*. *J. Gen. Microbiol.* 96, 165–174. doi: 10.1099/00221287-96-1-165

Jennings, M. L., and Cui, J. (2008). Chloride homeostasis in *Saccharomyces cerevisiae*: high affinity influx, V-ATPase-dependent sequestration, and identification of a candidate $Cl^−$ sensor. *J. Gen. Physiol.* 131, 379–391. doi: 10.1085/jgp.200709905

Jimenez, T., McDermott, J. P., Sánchez, G., and Blanco, G. (2011). Na,K-ATPase α4 isoform is essential for sperm fertility. *Proc. Natl. Acad. Sci. U.S.A.* 108, 644–649. doi: 10.1073/pnas.1016902108

Johansson, I., and Blatt, M. R. (2006). Interactive domains between pore loops of the yeast K+ channel TOK1 associate with extracellular $K^+$ sensitivity. *Biochem. J.* 393, 645–655. doi: 10.1042/BJ20051380

Johnson, B. R., Wyttenbach, R. A., Wayne, R., and Hoy, R. R. (2002). Action potentials in a giant algal cell: a comparative approach to mechanisms and evolution of excitability. *J. Undergrad. Neurosci. Educ.* 1, A23–A27.

Joubert, O., Nehmé, R., Bidet, M., and Mus-Veteau, I. (2010). Heterologous expression of human membrane receptors in the yeast *Saccharomyces cerevisiae*. *Methods Mol. Biol.* 601, 87–103. doi: 10.1007/978-1-60761-344-2_6 Kahm, M. (2011). *A Mathematical Model of the Potassium Homeostasis in Saccharomyces cerevisiae.* Dissertation zur Erlangung des Doktorgrades (Dr. rer. nat.) der Mathematisch-Naturwissenschaftlichen Fakultät der Rheinischen Friedrich-Wilhelms-Universität Bonn, 94. Available online at: http://hss.ulb.uni-bonn.de/2012/2903/2903.pdf

Kaji, D. M. (1993). Effect of membrane potential on K-Cl transport in human erythrocytes. *Am. J. Physiol.* 264, C376–C382.

Kanzaki, M., Nagasawa, M., Kojima, I., Sato, C., Naruse, K., Sokabe, M., et al. (1999). Molecular identification of a eukaryotic, stretch-activated nonselective cation channel. *Science* 285, 882–886. doi: 10.1126/science.285.5429.882

Ketchum, K. A., Joiner, W. J., Sellers, A. J., Kaczmarek, L. K., and Goldstein, S. A. (1995). A new family of outwardly rectifying potassium channel proteins with two pore domains in tandem. *Nature* 376, 690–695. doi: 10.1038/376690a0

Kinclová, O., Ramos, J., Potier, S., and Sychrová, H. (2001). Functional study of the *Saccharomyces cerevisiae* Nha1p C-terminus. *Mol. Microbiol.* 40, 656–668. doi: 10.1046/j.1365-2958.2001.02412.x

Kinclova-Zimmermannova, O., Gaskova, D., and Sychrova, H. (2006). The $Na^+,K^+/H^+$-antiporter Nha1 influences the plasma membrane potential of *Saccharomyces cerevisiae*. *FEMS Yeast Res.* 6, 792–800. doi: 10.1111/j.15671364.2006.00062.x

Kinclova-Zimmermannova, O., Zavrel, M., and Sychrova, H. (2005). Identification of conserved prolyl residue important for transport activity and the substrate specificity range of yeast plasma membrane Na$^+$/H$^+$ antiporters. *J. Biol. Chem.* 280, 30638–30647. doi: 10.1074/jbc.M506341200

Ko, C. H., and Gaber, R. F. (1991). TRK1 and TRK2 encode structurally related K$^+$ transporters in *Saccharomyces cerevisiae*. *Mol. Cell. Biol.* 11, 4266–4273.

Kolacna, L., Zimmermannova, O., Hasenbrink, G., Schwarzer, S., Ludwig, J., Lichtenberg-Fraté, H., et al. (2005). New phenotypes of functional expression of the mKir2.1 channel in potassium efflux-deficient *Saccharomyces cerevisiae* strains. *Yeast* 22, 1315–1323. doi: 10.1002/yea.1333

Kole, M. H., Ilschner, S. U., Kampa, B. M., Williams, S. R., Ruben, P. C., and Stuart, G. J. (2008). Action potential generation requires a high sodium channel density in the axon initial segment. *Nat. Neurosci.* 11, 178–186. doi: 10.1038/nn2040

Kung, C., Martinac, B., and Sukharev, S. (2010). Mechanosensitive channels in microbes. *Annu. Rev. Microbiol.* 64, 313–329. doi: 10.1146/annurev.micro.112408.134106

Kuroda, T., Bihler, H., Bashi, E., Slayman, C. L., and Rivetta, A. (2004). Chloride channel function in the yeast TRK-potassium transporters. *J. Membr. Biol.* 198, 177–192. doi: 10.1007/s00232-004-0671-1

Lecar, H., Ehrenstein, G., Binstock, L., and Taylor, R. E. (1967). Removal of potassium negative resistance in perfused squid giant axons. *J. Gen. Physiol.* 50, 1499–1515. doi: 10.1085/jgp.50.6.1499

Lee, S., Tong, L., and Denu, J. M. (2008). Quantification of endogenous sirtuin metabolite O-acetyl-ADP-ribose. *Anal. Biochem.* 383, 174–179. doi: 10.1016/j.ab.2008.08.033

Lemaire, T., Kaiser, J., Naili, S., and Sansalone, V. (2010). Modelling of the transport in electrically charged porous media including ionic exchanges. *Mech. Res. Commun.* 37, 495–499. doi: 10.1016/j.mechrescom.2010.05.009 Lesage, F., Guillemare, E., Fink, M., Duprat, F., Lazdunski, M., Romey, G., et al. (1996). A pH-sensitive yeast outward rectifier K$^+$ channel with two pore domains and novel gating properties. *J. Biol. Chem.* 271, 4183–4187. doi: 10.1074/jbc.271.8.4183

Levitan, I. B., and Kaczmarek, L. K. (2001). *The Neuron: Cell and Molecular Biology, 3rd Edn.* New York, NY; Oxford: Oxford University Press, 632.

Lichtenberg, H. C., Giebeler, H., and Höfer, M. (1988). Measurements of electrical potential differences across yeast plasma membranes with microelectrodes are consistent with values from steady-state distribution of tetraphenylphosphonium in *Pichia humboldtii*. *J. Membr. Biol.* 103, 255–261. doi: 10.1007/BF01993985

Liebeskind, B. J., Hillis, D. M., and Zakon, H. H. (2011). Evolution of sodium channels predates the origin of nervous systems in animals. *Proc. Natl. Acad. Sci. U.S.A.* 108, 9154–9159. doi: 10.1073/pnas.1106363108

Lipke, P. N., and Ovalle, R. (1998). Cell wall architecture in yeast: new structure and new challenges. *J. Bacteriol.* 180, 3735–3740.

Loret, M. O., Pedersen, L., and François, J. (2007). Revised procedures for yeast metabolites extraction: application to a glucose pulse to carbon-limited yeast cultures, which reveals a transient activation of the purine salvage pathway. *Yeast* 24, 47–60. doi: 10.1002/yea.1435

Loukin, S. H., and Saimi, Y. (1999). K$^+$-dependent composite gating of the yeast K$^+$ channel, Tok1. *Biophys. J.* 77, 3060–3070. doi: 10.1016/S00063495(99)77137-7

Loukin, S. H., Vaillant, B., Zhou, X. L., Spalding, E. P., Kung, C., and Saimi, Y. (1997). Random mutagenesis reveals a region important for gating of the yeast K$^+$ channel Ykc1. *EMBO J.* 16, 4817–4825. doi: 10.1093/emboj/16.16.4817 Madrid, R., Gómez, M. J., Ramos, J., and Rodríguez-Navarro, A. (1998). Ectopic potassium uptake in trk1 trk2 mutants of *Saccharomyces cerevisiae* correlates with a highly hyperpolarized membrane potential. *J. Biol. Chem.* 273, 14838–14844. doi: 10.1074/jbc.273.24.14838

Malagoli, P., Britto, D. T., Schulze, L. M., and Kronzucker, H. J. (2008). Futile Na$^+$ cycling at the root plasma membrane in rice (*Oryza sativa* L.): kinetics, energetics, and relationship to salinity tolerance. *J. Exp. Bot.* 59, 4109–4117. doi: 10.1093/jxb/ern249

Malínská, K., Malínski, J., Opekarová, M., and Tanner, W. (2003). Visualization of protein compartmentation within the plasma membrane of living yeast cells. *Mol. Biol. Cell* 14, 4427–4436. doi: 10.1091/mbc.E03-04-0221

Malinsky, J., Opekarová, M., and Tanner, W. (2010). The lateral compartmentation of the yeast plasma membrane. *Yeast* 27, 473–478. doi: 10.1002/yea.1772

Maresova, L., Urbankova, E., Gaskova, D., and Sychrova, H. (2006). Measurements of plasma membrane potential changes in *Saccharomyces cerevisiae* cells reveal the importance of the Tok1 channel in membrane potential maintenance. *FEMS Yeast Res.* 6, 1039–1046. doi: 10.1111/j.1567-1364.2006.00140.x

Márquez, J. A., and Serrano, R. (1996). Multiple transduction pathways regulate the sodium-extrusion gene PMR2/ENA1 during salt stress in yeast. *FEBS Lett.* 382, 89–92. doi: 10.1016/0014-5793(96)00157-3

Mäser, P., Hosoo, Y., Goshima, S., Horie, T., Eckelman, B., Yamada, K., et al. (2002). Glycine residues in potassium channel-like selectivity filters determine potassium selectivity in four-loop-per-subunit HKT transporters from plants. *Proc. Natl. Acad. Sci. U.S.A.* 99, 6428–6433. doi: 10.1073/pnas.082123799

Meychik, N. R., Honarmand, S. J., and Yermakov, I. P. (2007). Organic cation diffusion in root cell walls of different plants. *Biologija* 53, 60–63.

Meychik, N. R., Nikolaeva, Y. I., and Yermakov, I. P. (2006). Ion-exchange properties of cell walls of *Spinacia oleracea* L. roots under different environmental salt conditions. *Biochem. Mosc.* 71, 781–789. doi: 10.1134/S000629790607011X

Meychik, N. R., and Yermakov, I. P. (2001a). Swelling of root cell walls as an indicator of their functional state. *Biochemistry (Mosc).* 66, 178–187. doi: 10.1023/A:1002843615188

Meychik, N. R., and Yermakov, I. P. (2001b). Ion exchange properties of plant root cell walls. *Plant Soil.* 234, 181–193. doi: 10.1023/A:1017936318435

Meychik, N. R., and Yermakov, I. P. (2001c). Ion-exchange properties of cell walls isolated from lupine roots. *Biochemistry (Mosc).* 66, 556–563. doi: 10.1023/A:1010219221077

Minc, N., Boudaoud, A., and Chang, F. (2009). Mechanical forces of fission yeast growth. *Curr. Biol.* 19, 1096–1101. doi: 10.1016/j.cub.2009.05.031

Miranda, M., Bashi, E., Vylkova, S., Edgerton, M., Slayman, C., and Rivetta, A. (2009). Conservation and dispersion of sequence and function in fungal TRK potassium transporters: focus on *Candida albicans*. *FEMS Yeast Res.* 9, 278–292.

doi: 10.1111/j.1567-1364.2008.00471.x

Mitsui, K., Hatakeyama, K., Matsushita, M., and Kanazawa, H. (2009). *Saccharomyces cerevisiae* Na$^+$/H$^+$ antiporter Nha1p associates with lipid rafts and requires sphingolipid for stable localization to the plasma membrane. *J. Biochem.* 145, 709–720. doi: 10.1093/jb/mvp032

Mulet, J. M., and Serrano, R. (2002). Simultaneous determination of potassium and rubidium content in yeast. *Yeast* 19, 1295–1298. doi: 10.1002/yea.909

Murguía, J. R., Bellés, J. M., and Serrano, R. (1996). The yeast HAL2 nucleotidase is an *in vivo* target of salt toxicity. *J. Biol. Chem.* 271, 29029–29033. doi: 10.1074/jbc.271.46.29029

Muzzey, D., Gómez-Uribe, C. A., Mettetal, J. T., and van Oudenaarden, A. (2009). A systems-level analysis of perfect adaptation in yeast osmoregulation. *Cell* 138, 160–171. doi: 10.1016/j.cell.2009.04.047

Navarrete, C., Petrezsélyová, S., Barreto, L., Martínez, J. L., Zahrádka, J., Ariño, J., et al. (2010). Lack of main K$^+$ uptake systems in *Saccharomyces cerevisiae* cells affects yeast cell physiological parameters both in potassium sufficient and limiting conditions. *FEMS Yeast Res.* 10, 508–517. doi: 10.1111/j.15671364.2010.00630.x

Navarro, B., Kirichok, Y., and Clapham, D. E. (2007). KSper, a pH-sensitive K$^+$ current that controls sperm membrane potential. *Proc. Natl. Acad. Sci. U.S.A.* 104, 7688–7692. doi: 10.1073/pnas.0702018104

Newsholme, E. A., and Crabtree, B. (1976). Substrate cycles in metabolic regulation and in heat generation. *Biochem. Soc. Symp.* 41, 61–109.

Nielsen, L. J., Olsen, L. F., and Ozalp, V. C. (2010). Aptamers embedded in polyacrylamide nanoparticles: a tool for *in Vivo* metabolite sensing. *ACS Nano* 4, 4361–4370. doi: 10.1021/nn100635j

Ohgaki, R., Nakamura, N., Mitsui, K., and Kanazawa, H. (2005). Characterization of the ion transport activity of the budding yeast Na+/H+ antiporter, Nha1p, using isolated secretory vesicles. *Biochim. Biophys. Acta* 1712, 185–196. doi: 10.1016/j.bbamem.2005.03.011

Österlund, T., Nookaew, I., Bordel, S., and Nielsen, J. (2013). Mapping conditiondependent regulation of metabolism in yeast through genome-scale modeling. *BMC Syst. Biol.* 7:36. doi: 10.1186/1752-0509-7-36

Ozalp, V. C., Pedersen, T. R., Nielsen, L. J., and Olsen, L. F. (2010). Time-resolved measurements of intracellular ATP in the yeast *Saccharomyces cerevisiae* using a new type of nanobiosensor. *J. Biol. Chem.* 285, 37579–37588. doi: 10.1074/jbc.M110.155119

Palmer, C., Zhou, X.-L., Lin, J., Loukin, S., Kung, C., and Saimi, Y. (2001). A TRP homolog in *Saccharomyces cerevisiae* forms an intracellular $Ca^{2+}$ permeable channel in the yeast vacuolar membrane. *Proc. Natl. Acad. Sci. U.S.A.* 98, 7801–7805. doi: 10.1073/pnas.141036198

Paulsen, I. T., Sliwinski, M. K., Nelissen, B., Goffeau, A., and Saier, M. Jr. (1998). Unified inventory of established and putative transporters encoded within the complete genome of *Saccharomyces cerevisiae*. *FEBS Lett.* 430, 116–125. doi: 10.1016/S0014-5793(98)00629-2

Peiter, E., Fischer, M., Sidaway, K., Roberts, S. K., and Sanders, D. (2005). The *Saccharomyces cerevisiae* $Ca^{2+}$ channel Cch1pMid1p is essential for tolerance to cold stress and iron toxicity. *FEBS Lett.* 579, 5697–5703. doi: 10.1016/j.febslet.2005.09.058

Peña, A., Sánchez, N. S., and Calahorra, M. (2010). Estimation of the electric plasma membrane potential difference in yeast with fluorescent dyes: comparative study of methods. *J. Bioenerg. Biomembr.* 42, 419–432. doi: 10.1007/s10863-010-9311-x

Perlin, D. S., Harris, S. L., Seto-Young, D., and Haber, J. E. (1989). Defective H+-ATPase of hygromycin B-resistant *pma1* mutants from *Saccharomyces cerevisiae*. *J. Biol. Chem.* 264, 21857–21864.

Permyakov, S., Suzina, N., and Valiakhmetov, A. (2012). Activation of H+ATPase of the plasma membrane of *Saccharomyces cerevisiae* by glucose: the role of sphingolipid and lateral enzyme mobility. *PLoS ONE* 7:e30966. doi: 10.1371/journal.pone.0030966

Plášek, J., Gášková, D., Lichtenberg-Fraté, H., Ludwig, J., and Höfer, M. (2012). Monitoring of real changes of plasma membrane potential by diS-$C_3$(3). fluorescence in yeast cell suspensions. *J. Bioenerg. Biomembr.* 44, 559–569. doi: 10.1007/s10863-012-9458-8

Plášek, J., Gášková, D., Ludwig, J., and Höfer, M. (2013). Early changes in membrane potential of *Saccharomyces cerevisiae* induced by varying extracellular $K^+$, $Na^+$ or $H^+$ concentrations. *J. Bioenerg. Biomembr.* 45, 561–568. doi: 10.1007/s10863-013-9528-6

Posas, F., Chambers, J. R., Heyman, J. A., Hoeffler, J. P., de Nadal, E., and Ariño, J. (2000). The transcriptional response of yeast to saline stress. *J. Biol. Chem.* 275, 17249–17255. doi: 10.1074/jbc.M910016199

Pradet, A., and Raymond, P. (1983). Adenine nucleotide ratios and adenylate energy charge in energy metabolism. *Annu. Rev. Plant Physiol.* 34, 199–224. doi: 10.1146/annurev.pp.34.060183.001215

Prior, C., Potier, S., Souciet, J. L., and Sychrova, H. (1996). Characterization of the *NHA1* gene encoding Na+/H+-antiporter of the yeast *Saccharomyces cerevisiae*. *FEBS Lett.* 387, 89–93. doi: 10.1016/0014-5793(96)00470-X

Qian, H., and Beard, D. A. (2006). Metabolic futile cycles and their functions: a systems analysis of energy and control. *IEE Proc. Syst. Biol.* 153, 192–200. doi: 10.1049/ip-syb:20050086

Ramirez, J. A., Vacata, V., McCusker, J. H., Haber, J. E., Mortimer, R. K., Owen, W. G., et al. (1989). ATP-sensitive K+ channels in a plasma membrane H+-ATPase mutant of the yeast *Saccharomyces cerevisiae*. *Proc. Natl. Acad. Sci. U.S.A.* 86, 7866–7870. doi: 10.1073/pnas.86.20.7866

Ren, D., Navarro, B., Xu, H., Yue, L., Shi, Q., and Clapham, D. E. (2001). A prokaryotic voltage-gated sodium channel. *Science* 294, 2372–2375. doi: 10.1126/science.1065635

Rink, T. J. (1977). Membrane potential of guinea-pig spermatozoa. *J. Reprod. Fertil.* 51, 155–157. doi: 10.1530/jrf.0.0510155

Ritchie, R. J., and Larkum, A. W. D.,(1982). Cation exchange properties of the cell walls of *Enteromorpha intestinalis* (L.) Link. *(Ulvales Chlorophyta). J. Exp. Bot.* 132, 125–139. doi: 10.1093/jxb/33.1.125

Rivetta, A., Kuroda, T., and Slayman, C. (2011). Anion currents in yeast K+ transporters (TRK) characterize a structural homologue of ligandgated ion channels. *Pflugers Arch.* 462, 315–330. doi: 10.1007/s00424-0110959-9

Rivetta, A., Slayman, C., and Kuroda, T. (2005). Quantitative modeling of chloride conductance in yeast TRK potassium transporters. *Biophys. J.* 89, 2412–2426. doi: 10.1529/biophysj.105.066712

Roberts, S. K., Fischer, M., Dixon, G. K., and Sanders, D. (1999). Divalent Cation Block of Inward Currents and Low-Affinity K+ Uptake in *Saccharomyces cerevisiae*. *J. Bacteriol.* 181, 291–297.

Rösch, P., Harz, M., Schmitt, M., and Popp, J. (2005). Raman spectroscopic identification of single yeast cells. *J. Raman Spectrosc.* 36, 377–379. doi: 10.1002/jrs.1312

Ryan, P. R., Newman, I. A., and Arif, I. (1992). Rapid calcium exchange for protons and potassium in cell walls of *Chara*. *Plant Cell Environ.* 15, 675–683. doi: 10.1111/j.1365-3040.1992.tb01009.x

Saito, T., Soga, K., Hoson, T., and Terashima, I. (2006). The bulk elastic modulus and the reversible properties of cell walls in developing *Quercus* leaves. *Plant Cell Physiol.* 47, 715–725. doi: 10.1093/pcp/pcj042

Salton, M. R. J., and Kim, K.-S. (1996). "Structure," in *Medical Microbiology, 4th Edn.*, ed S. Baron (Galveston, TX: University of Texas Medical Branch at Galveston), Chapter 2.

Samoilov, M., Plyasunov, S., and Arkin, A. (2005). Stochastic amplification and signaling in enzymatic futile cycles through noise-induced bistability with oscillations. *Proc. Natl. Acad. Sci. U.S.A.* 102, 2310–2315. doi: 10.1073/pnas.0406841102

Satoh, H., Delbridge, L. M., Blatter, L. A., and Bers, D. M. (1996). Surface:volume relationship in cardiac myocytes studied with confocal microscopy and membrane capacitance measurements: species-dependence and developmental effects. *Biophys. J.* 70, 1494–1504. doi: 10.1016/S0006-3495(96)79711-4

Schaber, J., Adrover, M. A., Eriksson, E., Pelet, S., Petelenz-Kurdziel, E., Klein, D., et al. (2010). Biophysical properties of *Saccharomyces cerevisiae* and their relationship with HOG pathway activation. *Eur. Biophys. J.* 39, 1547–1556. doi: 10.1007/s00249-010-0612-0

Schwarzer, S., Kolacna, L., Lichtenberg-Fraté, H., Sychrova, H., and Ludwig, J. (2008). Functional expression of the voltage-gated neuronal mammalian potassium channel rat *ether à go-go1* in yeast. *FEMS Yeast Res.* 8, 405–413. doi: 10.1111/j.1567-1364.2007.00351.x

Sentenac, H., and Grignon, C. (1981). A model for predicting ionic equilibrium concentrations in cell walls. *Plant Physiol.* 68, 415–419. doi: 10.1104/pp.68.2.415

Serrano, R. (1988). Structure and function of proton translocating ATPase in plasma membranes of plants and fungi. *Biochim. Biophys. Acta* 947, 1–28. doi: 10.1016/0304-4157(88)90017-2

Serrano, R. (1996). Salt tolerance in plants and microorganisms: toxicity targets and defense mechanisms. *Int. Rev. Cytol.* 165, 1–52.

Shabala, L., Ross, T., McMeekin, T., and Shabala, S. (2006). Non-invasive microelectrode ion flux measurements to study adaptive responses of microorganisms to the environment. *FEMS Microbiol. Rev.* 30, 472–486. doi: 10.1111/j.1574-6976.2006.00019.x

Shabala, S., and Newman, I. (2000). Salinity effects on the activity of plasma membrane H+ and $Ca^{2+}$ transporters in bean leaf mesophyll: masking role of the cell wall. *Ann. Bot.* 85, 681–686. doi: 10.1006/anbo.2000.1131

Sievernich, A., Wildt, L., and Lichtenberg-Fraté, H. (2004). *In vitro* bioactivity of 17α-estradiol. *J. Steroid Biochem. Mol. Biol.* 92, 455–463. doi: 10.1016/j.jsbmb.2004.09.004

Skou, J. C. (1998). Nobel lecture. The identification of the sodium pump. *Biosci. Rep.* 18, 155–169. doi: 10.1023/A:1020196612909

Slayman, C. L., and Slayman, C. W. (1962). Measurement of membrane potentials in Neurospora. *Science* 136, 876–877. doi: 10.1126/science.136.3519.876

Smith, A. E., Zhang, Z., Thomas, C. R., Kennith, E., Moxham, K. E., and Middelberg, A. P. J. (2000). The mechanical properties of *Saccharomyces cerevisiae*. *Proc. Natl. Acad. Sci. U.S.A.* 97, 9871–9874. doi: 10.1073/pnas.97.18.9871

Soltanian, S., Dhont, J., Sorgeloos, P., and Bossier, P. (2007). Influence of different yeast cell-wall mutants on performance and protection against pathogenic bacteria (*Vibrio campbellii*) in gnotobiotically-grown *Artemia*. *Fish Shellfish Immunol.* 23, 141–153. doi: 10.1016/j.fsi.2006.09.013

Spira, F., Mueller, N. S., Beck, G., von Olshausen, P., Beig, J., and WedlichSöldner, R. (2012). Patchwork organization of the yeast plasma membrane into numerous coexisting domains. *Nature Cell Biol.* 14, 640–648. doi: 10.1038/ncb2487

Stagljar, I., Korostensky, C., Johnsson, N., and te Heesen, S. (1998). A genetic system based on split-ubiquitin for the analysis of interactions between membrane proteins *in vivo*. *Proc. Natl. Acad. Sci. U.S.A.* 95, 5187–5192. doi: 10.1073/pnas.95.9.5187

Stenson, J. D. (2009). *Investigating the Mechanical Properties of Yeast Cells*. Ph.D. thesis, University of Birmingham, 253. Available online at: http://etheses.bham.ac.uk/304/1/Stenson09PhD.pdf

Stenson, J. D., Hartley, P., Wang, C., and Thomas, C. R. (2011). Determining the Mechanical Properties of Yeast Cell Walls. *Biotechnol. Prog.* 27, 505–512. doi: 10.1002/btpr.554

Steudle, E. (1993). "Pressure probe techniques: basic principles and application to studies of water and solute relations at the cell, tissue and organ level," in *Water Deficits: Plant Responses from Cell to Community*, eds J. A. C. Smith and H. Griffiths (Oxford: Bios Scientific Publishers), 5–36.

Sutak, R., Botebol, H., Blaiseau, P. L., Léger, T., Bouget, F. Y., Camadro, J. M., et al. (2012). A comparative study of iron uptake mechanisms in marine microalgae: iron binding at the cell surface is a critical step. *Plant Physiol.* 160, 2271–2284. doi: 10.1104/pp.112.204156

Taglicht, D., Padan, E., and Schuldiner, S. (1991). Overproduction and purification of a functional $Na^+/H^+$ antiporter coded by nhaA (ant) from *Escherichia coli*. *J. Biol. Chem.* 266, 11289–11294.

Takami, H., and Horikoshi, K. (1999). Reidentification of facultatively alkaliphilic Bacillus sp. C-125 to Bacillus halodurans. *Biosci. Biotechnol. Biochem.* 63, 943–945. doi: 10.1271/bbb.63.943

Tálos, K., Pernyeszi, T., Majdik, C., Hegedûsova, A., and Páger, C. (2012). Cadmium biosorption by baker's yeast in aqueous suspension. *J. Serb. Chem. Soc.* 77, 549–561. doi: 10.2298/JSC110520181T

Taylor, A. R. (2009). A fast $Na^+/Ca^{2+}$-based action potential in a marine diatom. *PLoS ONE* 4:e4966. doi: 10.1371/journal.pone.0004966

Teng, J., Goto, R., Iida, K., Kojima, I., and Iida, H. (2008). Ion-channel blocker sensitivity of voltage-gated calcium-channel homologue Cch1 in *Saccharomyces cerevisiae*. *Microbiology* 154, 3775–3781. doi: 10.1099/mic.0.2008/021089-0

Teng, J., Iida, K., Imai, A., Nakano, M., Tada, T., and Iida, H. (2013). Hyperactive and hypoactive mutations in Cch1, a yeast homologue of the voltagegated calcium-channel pore-forming subunit. *Microbiology* 159, 970–979. doi: 10.1099/mic.0.064030-0

Thaminy, S., Miller, J., and Stagljar, I. (2004). The split-ubiquitin membrane-based yeast two-hybrid system. *Methods Mol. Biol.* 261, 297–312. doi: 10.1385/159259-762-9:297

Thomas, S. L., Bouyer, G., Cueff, A., Egée, S., Glogowska, E., and Ollivaux, C. (2011). Ion channels in human red blood cell membrane: actors or relics? *Blood Cells Mol. Dis.* 46, 261–265. doi: 10.1016/j.bcmd.2011.02.007

Thonart, P., Custinne, M., and Paquot, M. (1982). Zeta potential of yeast cells: application in cell immobilization. *Enzyme Microb. Technol.* 4, 191–194. doi: 10.1016/0141-0229(82)90116-8

Tolla, D. A., Kiley, P. J., Lomnitz, J. G., and Savageau, M. A. (2015). Design principles of a conditional futile cycle exploited for regulation. *Mol. Biosyst*. doi: 10.1039/c5mb00055f. [Epub ahead of print].

Ton, V. K., and Rao, R. (2004). Functional expression of heterologous proteins in yeast: insights into $Ca^{2+}$ signaling and $Ca^{2+}$-transporting ATPases. *Am. J. Physiol. Cell Physiol.* 287, C580–C589. doi: 10.1152/ajpcell.00135.2004

Tyree, M. T. (1968). Determination of transport constants of isolated *Nitella* cell walls. *Can. J. Bot.* 46, 317–327. doi: 10.1139/b68-054

Ullah, A., Chandrasekaran, G., Brul, S., and Smits, G. J. (2013). Yeast adaptation to weak acids prevents futile energy expenditure. *Front. Microbiol.* 4:142. doi: 10.3389/fmicb.2013.00142

Vacata, V., Kotyk, A., and Sigler, K. (1981). Membrane potentials in yeast cells measured by direct and indirect methods. *Biochim. Biophys. Acta* 643, 265–268. doi: 10.1016/0005-2736(81)90241-8

Vallejo, C. G., and Serrano, R. (1989). Physiology of mutants with reduced expression of plasma membrane $H^+$-ATPase. *Yeast* 5, 307–319. doi: 10.1002/yea.320050411

Van Belle, D., and André, B. (2001). A genomic view of yeast membrane transporters. *Curr. Opin. Cell Biol.* 13, 389–398. doi: 10.1016/S09550674(00)00226-X

Volkmer, B., and Heinemann, M. (2011). Condition-dependent cell volume and concentration of Escherichia coli to facilitate data conversion for systems biology modeling. *PLoS ONE* 6:e23126. doi: 10.1371/journal.pone.0023126

Volkov, V., Boscari, A., Clement, M., Miller, A. J., Amtmann, A., and Fricke, W. (2009). Electrophysiological characterization of pathways for $K^+$ uptake into growing and non-growing leaf cells of barley. *Plant Cell Environ.* 32, 1778–1790. doi: 10.1111/j.1365-3040.2009.02034.x

Volkov, V., Hachez, C., Moshelion, M., Draye, X., Chaumont, F., and Fricke, W. (2007). Osmotic water permeability differs between growing and non-growing barley leaf tissues. *J. Exp. Bot.* 58, 377–390. doi: 10.1093/jxb/erl203

Walsby, A. E., Hayes, P. K., and Boje, R. (1995). The gas vesicles, buoyancy and vertical distribution of cyanobacteria in the Baltic Sea. *Eur. J. Phycol.* 30, 87–94. doi: 10.1080/09670269500650851

Waters, S., Gilliham, M., and Hrmova, M. (2013). Plant high-affinity potassium (HKT) transporters involved in salinity tolerance: structural insights to probe differences in ion selectivity. *Int. J. Mol. Sci.* 14, 7660–7680. doi: 10.3390/ijms14047660

Wei, C., and Lintilhac, P. M. (2007). Loss of stability: a new look at the physics of cell wall behavior during plant cell growth. *Plant Physiol.* 145, 763–772. doi: 10.1104/pp.107.101964

Widschwendter, M., Lichtenberg-Fraté, H., Hasenbrink, G., Schwarzer, S., Dawnay, A., Lam, A., et al. (2009). Serum oestrogen receptor a and b bioactivity are independently associated with breast cancer: a proof of principle study. *Br. J. Cancer.* 101, 160–165. doi: 10.1038/sj.bjc.6605106

Wieland, J., Nitsche, A. M., Strayle, J., Steiner, H., and Rudolph, H. K. (1995). The PMR2 gene cluster encodes functionally distinct isoforms of a putative $Na^+$ pump in the yeast plasma membrane. *EMBO J.* 14, 3870–3882.

Wilson, W. A., Hawley, S. A., and Hardie, D. G. (1996). Glucose repression/derepression in budding yeast: SNF1 protein kinase is activated by phosphorylation under derepressing conditions, and this correlates with a high AMP:ATP ratio. *Curr. Biol.* 6, 1426–1434. doi: 10.1016/S0960-9822(96)00747-6

Zahrádka, J., and Sychrova, H. (2012). Plasma-membrane hyperpolarization diminishes the cation efflux via Nha1 antiporter and Ena ATPase under potassium-limiting conditions. *FEMS Yeast Res.* 12, 439–446. doi: 10.1111/j.1567-1364.2012.00793.x

Zayats, V., Stockner, T., Pandey, S. K., Wörz, K., Ettrich, R., and Ludwig, J. (2015). A refined atomic scale model of the *Saccharomyces cerevisiae* $K^+$-translocation protein Trk1p combined with experimental evidence confirms the role of selectivity filter glycines and other key residues. *Biochim. Biophys. Acta* 1848, 1183–1195. doi: 10.1016/j.bbamem.2015.02.007

Zhou, X. L., Batiza, A. F., Loukin, S. H., Palmer, C. P., Kung, C., and Saimi, Y. (2003). The transient receptor potential channel on the yeast vacuole is mechanosensitive. *Proc. Natl. Acad. Sci. U.S.A.* 100, 7105–7110. doi: 10.1073/pnas.1230540100

**Conflict of Interest Statement:** The author declares that the research was conducted in the absence of any commercial or financial relationships that could be construed as a potential conflict of interest.